\documentstyle{mn}
\newif\ifAMStwofonts
\AMStwofontstrue

\input{epsf}

\def\ebv{E_{B-V}}
\def\tbb{kT_{\rm bb}}
\def\tin{kT_{\rm tr}}
\def\ee{$e^\pm$}
\def\g{$\gamma$}
\def\nh{N_{\rm H}}
\def\af{A_{\rm Fe}}

\def\ginga{{\it Ginga}}
\def\asca{{\it ASCA}}

\def\efe{E_{\rm Fe}}
\def\sfe{\sigma_{\rm Fe}}
\def\wfe{W_{\rm Fe}}
\def\ife{I_{\rm Fe}}

\def\mmax{\dot M_{\rm max}}

\ifoldfss
  \ifCUPmtlplainloaded \else
    \NewTextAlphabet{textbfit} {cmbxti10} {}
    \NewTextAlphabet{textbfss} {cmssbx10} {}
    \NewMathAlphabet{mathbfit} {cmbxti10} {} % for math mode
    \NewMathAlphabet{mathbfss} {cmssbx10} {} %  "   "    "
  \fi
  \ifAMStwofonts
    \ifCUPmtlplainloaded \else
      \NewSymbolFont{upmath} {eurm10}
      \NewSymbolFont{AMSa} {msam10}
      \NewMathSymbol{\upi}     {0}{upmath}{19}
      \NewMathSymbol{\umu}     {0}{upmath}{16}
      \NewMathSymbol{\upartial}{0}{upmath}{40}
      \NewMathSymbol{\leqslant}{3}{AMSa}{36}
      \NewMathSymbol{\geqslant}{3}{AMSa}{3E}

      \let\leq=\leqslant 
      \let\geq=\geqslant 
    \fi
  \fi
\fi % End of OFSS

\ifnfssone
  \newmathalphabet{\mathit}
  \addtoversion{normal}{\mathit}{cmr}{m}{it}
  \addtoversion{bold}{\mathit}{cmr}{bx}{it}
  \def\textbfit{\protect\txtbfit}
  
  \long\def\txtbfit#1{{\fontfamily{cmr}\fontseries{bx}\fontshape{it}%
    \selectfont #1}}

  \newmathalphabet{\mathbfit} % math mode version of \textbfit{..}
  \addtoversion{normal}{\mathbfit}{cmr}{bx}{it}
  \addtoversion{bold}{\mathbfit}{cmr}{bx}{it}
  \newmathalphabet{\mathbfss} % math mode version of \textbfss{..}
  \addtoversion{normal}{\mathbfss}{cmss}{bx}{n}
  \addtoversion{bold}{\mathbfss}{cmss}{bx}{n}
  \ifAMStwofonts
    \ifCUPmtlplainloaded \else
      \UseAMStwoboldmath
      \makeatletter
      \new@mathgroup\upmath@group
      \define@mathgroup\mv@normal\upmath@group{eur}{m}{n}
      \define@mathgroup\mv@bold\upmath@group{eur}{b}{n}
      \edef\UPM{\hexnumber\upmath@group}
      \new@mathgroup\amsa@group
      \define@mathgroup\mv@normal\amsa@group{msa}{m}{n}
      \define@mathgroup\mv@bold\amsa@group{msa}{m}{n}
      \edef\AMSa{\hexnumber\amsa@group}
      \makeatother
      \mathchardef\upi="0\UPM19
      \mathchardef\umu="0\UPM16
      \mathchardef\upartial="0\UPM40
      \mathchardef\leqslant="3\AMSa36
      \mathchardef\geqslant="3\AMSa3E

      \let\leq=\leqslant 
      \let\geq=\geqslant 
    \fi
  \fi
\fi % End of NFSS release 1

\ifnfsstwo
  \def\textbfit{\protect\txtbfit}
  
  \long\def\txtbfit#1{{\fontfamily{cmr}\fontseries{bx}\fontshape{it}%
    \selectfont #1}}

  \DeclareMathAlphabet{\mathbfit}{OT1}{cmr}{bx}{it}
  \SetMathAlphabet\mathbfit{bold}{OT1}{cmr}{bx}{it}
  \DeclareMathAlphabet{\mathbfss}{OT1}{cmss}{bx}{n}
  \SetMathAlphabet\mathbfss{bold}{OT1}{cmss}{bx}{n}
  \ifAMStwofonts
    \ifCUPmtlplainloaded \else
      \DeclareSymbolFont{UPM}{U}{eur}{m}{n}
      \SetSymbolFont{UPM}{bold}{U}{eur}{b}{n}
      \DeclareSymbolFont{AMSa}{U}{msa}{m}{n}
      \DeclareMathSymbol{\upi}{0}{UPM}{"19}
      \DeclareMathSymbol{\umu}{0}{UPM}{"16}
      \DeclareMathSymbol{\upartial}{0}{UPM}{"40}
      \DeclareMathSymbol{\leqslant}{3}{AMSa}{"36}
      \DeclareMathSymbol{\geqslant}{3}{AMSa}{"3E}

      \let\leq=\leqslant 
      \let\geq=\geqslant 
    \fi
  \fi
\fi % End of NFSS release 2

\ifCUPmtlplainloaded \else
  \ifAMStwofonts \else % If no AMS fonts
    \def\upi{\pi}
    \def\umu{\mu}
    \def\upartial{\partial}
  \fi
\fi

\title[X-ray/$\gamma$-ray spectra and binary parameters of GX 339--4]
{Broad-band X-ray/$\bmath{\gamma}$-ray spectra and binary parameters\\
of GX 339--4 and their astrophysical implications}

\author[A. A. Zdziarski et al.]
{\parbox[]{7in} {Andrzej A. Zdziarski$^1$, Juri Poutanen$^2$, Joanna
Miko{\l}ajewska$^1$,\\ Marek Gierli\'nski$^{3,1}$, Ken Ebisawa$^4$, and W. Neil
Johnson$^5$ }\\
 $^1$N. Copernicus Astronomical Center, Bartycka 18, 00-716 Warsaw, Poland \\
$^2$Stockholm  Observatory, S-133 36 Saltsj\"obaden, Sweden \\
$^3$Astronomical Observatory, Jagiellonian University, Orla 171, 30-244
Cracow, Poland \\
$^4$Laboratory for High Energy Astrophysics,
NASA/Goddard Space Flight Center, Greenbelt, MD 20771, USA \\
$^5$E. O. Hulburt Center for Space Research,
Naval Research Laboratory, Washington, DC 20375, USA \\
}

\date{Accepted 1998 July 28. Received 1998 January 2}

\pagerange{\pageref{firstpage}--\pageref{lastpage}}
\pubyear{1998}

\begin{document}

\maketitle

\label{firstpage}

\begin{abstract} We present  X-ray/\g-ray spectra of the binary GX 339--4
observed in the hard state simultaneously by \ginga\/ and {\it CGRO\/} OSSE
during an outburst in 1991 September. The \ginga\/ X-ray spectra are well
represented by a power law with a photon spectral index of $\Gamma\simeq 1.75$
and a Compton reflection component with a fluorescent Fe K$\alpha$ line
corresponding to a solid angle of an optically-thick, ionized, medium of $\sim
0.4\times 2\pi$. The OSSE data ($\geq 50$ keV) require a sharp high-energy
cutoff in the power-law spectrum. The broad-band spectra are very well modelled
by repeated Compton scattering in a thermal plasma with an optical depth of
$\tau\sim 1$ and $kT\simeq 50$ keV. We also study the distance to the system
and find it to be $\ga 3$ kpc, ruling out earlier determinations of $\sim 1$
kpc. Using this limit, the observed reddening and the orbital period, we find
the allowed range of the mass of the primary is consistent with it being a
black hole.

We find the data are incosistent with models of either homogenous or patchy
coronae above the surface of an accretion disc. Rather, they are consistent
with the presence of a hot inner hot disc with the viscosity parameter of
$\alpha\sim 1$ accreting at a rate close to the maximum set by advection. The
hot disc is surrounded by a cold outer disc, which gives rise to the reflection
component and a soft X-ray excess, also present in the data. The seed photons
for Comptonization are unlikely to be due to thermal synchrotron radiation.
Rather, they are supplied by the outer cold disc and/or cold clouds within the
hot disc. \ee\ pair production is negligible if electrons are thermal. The hot
disc model, which scaled parameters are independent of the black-hole mass, is
supported by the similarity of the spectrum of GX 339--4 to those of other
black-hole binaries and Seyfert 1s. On the other hand, their spectra in the
soft \g-ray regime are significantly harder than those of weakly-magnetized
neutron stars. Based on this difference, we propose that the presence of
broad-band spectra corresponding to thermal Comptonization with $kT\ga 50$ keV
represents a black-hole signature.

\end{abstract}


\section{INTRODUCTION}
\label{s:intro}

GX 339--4, a bright and well-studied binary X-ray source, is commonly
classified as a black hole candidate based on the similarity of its X-ray
spectral states and short-time variability to those of Cyg X-1 (e.g.\ Tanaka \&
Lewin 1995). However, determinations of the mass of its compact star, $M_{\rm
X}$, have been inconclusive (e.g.\ Cowley, Crampton \& Hutchings 1987,
hereafter C87; Callanan et al.\ 1992, hereafter C92), and thus its nature has
been uncertain. Therefore, further studies of the properties of GX 339--4 as
well as their comparison to those of objects with more direct evidence for
harbouring a black hole is of crucial importance.

In this work, we present two, very similar, broad-band X-ray/\g-ray (hereafter
X\g) spectra of GX 339--4 obtained during a strong outburst of the source in
September 1991 (Harmon et al.\ 1994) simultaneously by \ginga\/ (Makino et al.\
1987) and the Oriented Scintillation Spectroscopy Experiment (OSSE) detector
(Johnson et al.\ 1993) on board the {\it Compton Gamma Ray Observatory\/} ({\it
CGRO}). The source was in the hard (also called `low') spectral state. The
\ginga\/ and OSSE observations were reported separately by Ueda, Ebisawa \&
Done (1994, hereafter U94) and Grabelsky et al.\ (1995, hereafter G95),
respectively. However, the data from the two instruments have not been fitted
together, and, e.g.\ G95 found models with Compton reflection of X\g\ photons
from an accretion disc unlikely whereas U94 found strong evidence in the
\ginga\/ data for the presence of this process.

Here, we re-analyze the simultaneous \ginga\/ and OSSE data based on the
present accurate calibration of those instruments. This leads to a
reconciliation of the apparent discrepancies between the data sets from the two
instruments, and allows us to fit the joint data with physical models. We also
study the distance, reddening, Galactic column density and the masses of the
binary members. Those results are then used in studying radiative processes,
geometry and physical models of the source. Finally, we find the X\g\ spectrum
of GX 339--4 similar to those of black-hole binaries and Seyfert AGNs, and, in
particular, virtually identical to that of NGC 4151, the Seyfert brightest in
hard X-rays. This favours physical models with scaled parameters independent of
the central mass, such as a hot accretion disc with unsaturated thermal
Comptonization (Shapiro, Lightman \& Eardley 1976, hereafter S76). On the other
hand, the spectrum of GX 339--4 is significantly different from those observed
from neutron star binaries, which supports the black-hole nature of the compact
object in GX 339--4.

\section{THE PARAMETERS OF THE BINARY}

In order to analyze the X-ray data meaningfully, we need to estimate basic
parameters of the binary system. Of importance here are the Galactic column
density, $\nh$, the interstellar reddening, $\ebv$, the distance, $d$ (for
which published estimates range from 1.3 to 4 kpc), the masses of the primary
and secondary, $M_{\rm X}$ and $M_{\rm c}$, respectively, and the inclination
(with respect to the normal to the orbital plane), $i$.

\subsection{Reddening and column density}

Grindlay (1979) found strong interstellar Na\,{\sc i} D absorption lines and
diffuse interstellar bands at $\lambda \sim  5775$--5795, 6010, 6176, 6284, and
6376 \AA, while C87 found a strong interstellar Ca\,{\sc ii} K absorption line
and diffuse $\lambda 4430$\, \AA\ absorption band. The equivalent widths of
these features are consistent with $\ebv \simeq 1$--1.3. From the uncertainties
of the published estimates, we derive the weighted mean of \begin{equation}
\ebv=1.2 \pm 0.1\,. \end{equation}

The most extended all-sky study of the distribution of neutral H based on
high-resolution {\it IUE\/} observations of Ly$\alpha$ absorption towards 554
OB stars shows their $\nh$ well correlated with the column density of dust,
measured by $\ebv$, with $\langle \nh/\ebv\rangle = 4.93 \times 10^{21}\, {\rm
cm^{-2}\, mag^{-1}}$ (Diplas \& Savage 1994). $\ebv = 1.2 \pm 0.1$ derived
above thus indicates \begin{equation} \nh = (6.0 \pm 0.6) \times 10^{21} \rm
cm^{-2}\,. \end{equation}

This $\nh$ is in excellent agreement with that derived from X-ray data. We
obtain $\nh=(6.2\pm 0.7) \times 10^{21}$ cm$^{-2}$ from the depth of the O edge
of $\tau_{\rm O}=2.6\pm 0.3$ measured by Vrtilek et al.\ (1991), and assuming
the O abundance of Anders \& Ebihara (1982). On the other hand, Vrtilek et al.\
(1991) and Ilovaisky et al.\ (1986) have obtained $\nh= (6.6\pm 0.3)\times
10^{21}$ cm$^{- 2}$ and $(5.0\pm 0.7)\times 10^{21}$ cm$^{- 2}$ from continuum
fits in the soft and hard state, respectively. Those values are less reliable
because those authors assumed the continuum models of optically-thin thermal
bremsstrahlung and a single power law for the corresponding two states. The
first model cannot hold for a luminous source based on the standard efficiency
argument, and the second model disagree with \ginga\/ data showing the presence
of a strong soft X-ray excess in addition to the harder power law over two
orders of magnitude of the flux in the hard state (U94). Therefore, we assume
hereafter $\nh=6\times 10^{21}$ cm$^{-2}$, which value agrees with both optical
and X-ray data.

On the other hand, Ilovaisky et al.\ (1986) obtained $\ebv=0.7\pm 0.1$ by
converting their fitted $\nh$ (which appears underestimated, see above) with an
old calibration of $\nh/\ebv$, which underestimates $\ebv(\nh)$ according to
the presently most extensive study of Diplas \& Savage (1994). This value
of $\ebv$ is also in disagreement with the results from interstellar features,
and it appears incorrect.

\subsection{Distance}

Obviously, reddening increases with distance. However, the resulting
correlation depends sensitively on direction because the distribution of the
interstellar matter is very complex and patchy (e.g.\ Diplas \& Savage 1994;
Neckel \& Klare 1980), especially towards the Galactic Center, in which
direction is the line of sight of GX 339--4.

To illustrate this problem, we show here examples of a wide dispersion of
$\ebv(d)$ with $l,b$ within $\pm 5\degr$ from the direction to GX 339--4. That
field includes many OB stars in stellar systems, for which both $\ebv$ and $d$
have been well studied. First, the distribution of OB stars and dust in a
21-deg$^2$ field centered at $(l,b)= (335\degr,0\degr$) was studied in detail
by FitzGerald (1987), who found the dust is distributed in two distinct clouds,
one in the local arm at $190 \pm 30$ pc and the other in an interarm cloud at
$690 \pm 70$ pc with $\langle\ebv\rangle = 0.21$ and 0.76, respectively. Three
OB associations along this line of sight found at $\langle d\rangle= 1.34$,
2.41 and 3.69 kpc have $\ebv=0.82$, 0.86 and 0.92, respectively. On the other
hand, the open clusters NGC 6200, NGC 6204 and Hogg 22 at $(l,b)\simeq
(338.5,-1.2)$ show $\ebv\leq 0.66$ for $d \leq 2.78$, while NGC 6193 at $(l,b)
= (336.7,-1.6)$ and NGC 6250 at (340.8,$- 1.8$) show $\ebv = 0.44$--0.49 at $d
= 1.4$ kpc, and $\ebv = 0.38$ at $d = 0.95$ kpc, respectively (Moffat \& Vogt
1973, 1975; FitzGerald et al.\ 1977; V\'azquez \& Feinstein 1992). Finally, the
clusters NGC 6208 at (333.5,$- 5.7$), IC 4651 at (341,$-8$) and NGC 6352 at
(342,$-7$) at $d=1$, 1.15 and 5 kpc show $\ebv= 0.18$, 0.12 and 0.4,
respectively (Lindoff 1972; Alcaino 1971). These results clearly show that
$\ebv(d)$ around GX 339--4 depends sensitively on direction.

\begin{figure} \begin{center} \leavevmode \epsfxsize=8cm \epsfbox{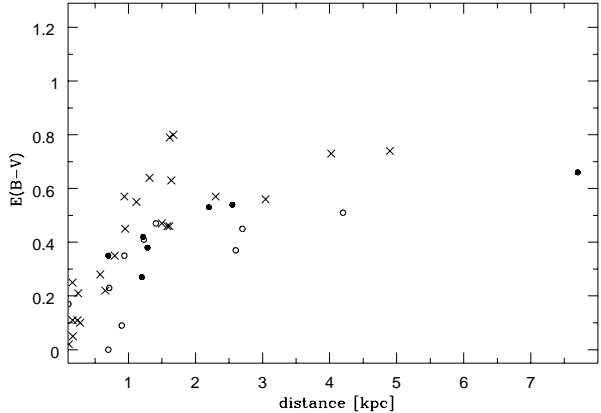}
\end{center} \caption{The dependence of $\ebv$ on distance in the field
centered at GX 339--4. Dots represent stars in a small field with a radius of
$\sim 1{\degr}$ around the source. Two stars closest to the line of sight of GX
339--4 are shown by the two dots between 2 and 3 kpc. Crosses and open circles
represent nearby sources outside that radius on the side of the Galactic Center
and on the opposite side, respectively. (Note an increase of extinction per
unit distance toward the Galactic Center.) On the other hand, the extinction of
GX 339--4 is $\ebv=1.2\pm 0.1$. } \end{figure}

Therefore, we have determined $\ebv$ and $d$ for $\sim 450$ stars with $V\la
10$ within $\pm 5\degr$ around GX 339--4 using their $UBV$ magnitudes and
spectral types from the catalogue of Mermilliod \& Mermilliod (1994). Our study
confirms the strong dependence of $\ebv(d)$ on direction and allows an accurate
determination of $\ebv(d,l,b)$ in that field. We have found that $\ebv$ depends
much more sensitively on $b$ than on $l$, and thus we present in Fig.\ 1 our
results only for a smaller field centered at GX 339--4 with ($l,b) = (338.9 \pm
5{\degr}, -4.3 \pm 1{\degr}$).  Stars with $l \la 338\degr$ and $\ga 340\degr$
are indicated by open circles and crosses, respectively. Dots represent stars
in a central field with a radius of $\sim 1{\degr}$. From Fig.\ 1, it is clear
that the extinction increases up to $\langle\ebv\rangle \approx 0.6$ at $d
\approx 2$ kpc. There is a possible further increase beyond $\sim 3$ kpc, but
to confirm it we would need a deeper survey than that of Mermilliod \&
Mermilliod (1994). Two stars nearest to GX 339--4 with known $UBV$ magnitudes
and spectral types are the early B-type giants HD\,153222 and
CPD\,$-48{\degr}11361$ at $(l,b) \approx (338{\degr},-4\degr)$, for which we
have found $\ebv = 0.53$, $d \ga 2.2$ kpc, and $\ebv = 0.54$, $d \simeq 2.6$
kpc, respectively. Thus, our strong conclusion is that GX 339--4 with the
measured $\ebv = 1.2 \pm 0.1$ must be located at $d> 2.6$ kpc.

This result rules out the most recent estimate of $d=1.3$ kpc towards GX
339--4, obtained from a {\it ROSAT\/} measurement of its X-ray halo by Predehl
et al.\ (1991). They suggest that the halo is produced by a dust cloud near the
source, and identify that cloud with the second of two dense clouds found at $d
\sim 100$--250 pc and $\sim 800$--1400 pc, respectively, by Neckel \& Klare
(1980). However, we have not been able to reproduce their identification. There
are indeed two clouds shown in the direction of $l=339\degr$ in Fig.\ 9 of
Neckel \& Klare (1980), but that figure gives the cloud distribution {\it
projected\/} onto the Galactic plane. In fact, Table IVa of Neckel \& Klare
(1980) shows that the field of GX 339--4 (numbered 196 in that paper) does not
contribute to extinction in those clouds, as well as Fig.\ 8c of that paper
shows that $A_V$ at 1 kpc in that field is $<1.9$ (whereas $A_V\sim 3.4$--4
corresponds to $\ebv=1.2\pm 0.1$ of GX 339--4). These findings are confirmed by
a more extensive study of Galactic extinction by FitzGerald (1987). She shows
the same clouds are located in the field of Ara OB 1 association at $(l,b)
\simeq (335\degr, 0\degr)$, confirming that they {\it do not\/} concide with
the line of sight to GX 339--4. Moreover, the total $\ebv \approx 0.8$ at $\sim
1.34$ kpc (FitzGerald 1987) produced by the two clouds is lower than the $\ebv
\simeq 1.2$ measured for GX 339--4 (and adopted by Predehl et al.\ 1991 in
their analysis). Similarly, $d=1.33$ kpc derived by Mauche \& Gorenstein (1986)
for GX 339--4 from {\it Einstein\/} observation of the X-ray halo corresponds
just to the Galaxy's dust layer, and thus it represents a lower limit to the
distance. That limit can be actually higher ($d \ga 2$ kpc) since recent
studies (Diplas \& Savage 1994) show a value of the scaleheight for the dust
layer of 150 pc that is larger than 100 pc adopted by Mauche \& Gorenstein.
Those two scaleheights predict $\ebv \sim 0.5$ and $\sim 0.8$, respectively,
both lower than the $\ebv$ of GX 339--4.

On the other hand,  the distance to GX 339--4 can be estimated from its
systemic velocity of $V_0 = -62 \pm 10$ km s$^{-1}$ (C92). Using the $V_0$-$d$
conversion chart from Burton (1992), the kinematic distance is $d = 4 \pm 1$
kpc. (Note that a possible peculiar velocity of the system as well as a
systematic uncertainty of the above $V_0$ makes this estimate less secure than
the extinction limit.)

%Parenthetically, we note that the radial velocity of $\sim -140$ km s$^{-1}$
%reported for the interstellar Ca\,{\sc ii} K by C87 is
%inconsistent with the galactic rotation for any reasonable distance. Namely,
%the radio observations of H\,{\sc ii} region complexes between $l =335.8\degr$
%and $l = 338.4\degr$ show radial velocities up to a lower velocity of $-93$ km
%s$^{-1}$, which are consistent with kinematic distances of $\sim 8$ kpc
%(Georgelin \& Georgelin 1976).

We note here that both the studies of distribution of matter in the Galaxy as
well as the measurements of radial velocities of H\,{\sc ii} regions in the
range $l \approx 300{\degr}$--$340{\degr}$ indicate the presence of a spiral
arm at a distance of $\sim 4$ kpc (Scheffler \& Elsasser 1987 and references
therein). Our derived kinematic distance and the relatively large $\ebv$ of GX
339--4 are consistent with its location in that arm.

Finally, we can put an upper limit on $d$ from the Eddington limit, which
appears to be satisfied in all known black-hole binaries (Tanaka \& Lewin
1995). The most luminous state of GX 339--4 reported as yet appears to be a
soft state in 1988 (Miyamoto et al.\ 1991), for which we estimate
the bolometric flux (using the disc blackbody model, see Section 5.2)
to be $\sim 4\times 10^{-8}$ erg cm$^{-2}$ s$^{-1}$.
This corresponds to $d\simeq
10 (L/L_{\rm E})^{1/2}
(M_{\rm X}/ 3 M_\odot)^{1/2}$ kpc, where $L_{\rm E}$ is the Eddington
luminosity (see Section 2.3 for discussion of $M_{\rm
X}$). On the other hand, typical maximum luminosities of black-hole binaries
are at least factor of a few below $L_{\rm E}$
(Tanaka \& Lewin 1995), and, e.g.\ $d\simeq 4.5$ kpc if $L=0.2L_{\rm E}$ and
$M_{\rm X}=3M_\odot$.

Concluding this section, our strongest limit is $d \ga 3$ kpc based on the
extinction study. This limit is consistent with the kinematic distance of $d =
4 \pm 1$ kpc and with the Eddington limit. Thus, we adopt $d=4$ kpc hereafter.

\subsection{The masses and geometry}

With the above results on the $\ebv$ and $d$, we can constrain the mass of the
companion star. In general, we expect $M_{\rm c}\la 1M_\odot$ in low-mass X-ray
binaries (hereafter LMXBs, e.g.\ van Paradijs \& McClintock 1995), to which
class GX 339--4 most likely belongs. At $d \simeq 4$ kpc and $A_V \sim 3.4$--4,
a $1 M_\odot$ main sequence star has an observed $V\simeq 20.5$--21 ($B-V
\simeq 1.8$, $V-R_{\rm c} \simeq 1$, $V-I_{\rm c} \simeq 2$), which corresponds
to fluxes below the faintest ones observed as yet from GX 339--4, as
illustrated in Fig.\ 2. Thus, the presence of a $1 M_\odot$ secondary star is
entirely consistent with the present data. We note that our conclusion differs
from that of C92, who claimed $M_{\rm c} \la 0.4M_\odot$. The reason for the
discrepancy is that those authors assumed $d=1.3$ kpc (from Predehl et al.\
1991) and $\ebv=0.7$ (from Ilovaisky et al.\ 1986), which both values have been
found by us to be incorrect, see above.

\begin{figure} \begin{center} \leavevmode \epsfxsize=4.5cm
\epsfbox{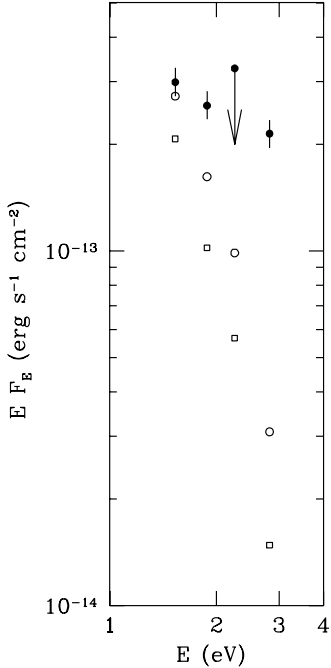} \end{center} \caption{Fluxes at the $I_c$, $R_c$, $V$,
and $B$ bands (from left to right) of a $1 M_\odot$ main sequence star at 4 kpc
reddened by $\ebv=1.2$ (open squares) and by $\ebv=1$ (a less likely value;
open circles) compared to the faintest observations (not simultaneous) of GX
339--4 (filled circles). The $B$, $V$ are from Ilovaisky \& Chevalier (1981)
and Ilovaisky et al.\ (1986), and $R_{\rm c}$, $I_{\rm c}$ are from C92. For
both estimates of $\ebv$, the stellar spectrum is below the observations of GX
339--4 (which spectrum is a sum of emission from the companion star and from
the accretion flow onto the compact object). Thus, the presence of a $1
M_\odot$ secondary star is fully compatible with the present data. }
\end{figure}

C92 detected a 14.8-h flux modulation in both optically-high and low states.
They also found the same periodicity in the radial velocities of He\,{\sc ii}
4686 and H\,{\sc i} Balmer emission lines measured in low resolution spectra
from 1986 May 7--9 (C87). We caution here that interpreting this periodicity as
due to the orbital period remains relatively uncertain.

First, the periodicity is absent in the radial velocity data from 1985 March
14--17 of C87, although it is possibly due to their poorer quality. Second,
although C87 claimed that the source was in its `normal' (high optically) state
of $\sim 16.5$ mag both in 1985 March and 1986 May, C92 noted a displacement
in the mean velocity between those data from $\sim +96$ km s$^{-1}$ to $-62 \pm
10$ km s$^{-1}$, as well as Ilovaisky et al.\ (1986) reported a relatively
faint optical state from late 1985 March ($B \sim 18.8$) through 1985 April 29,
which contradict this claim of C87. Third, Corbet et al.\ (1987) shows
strong variability around 1986 May. Namely, their optical photometry from
1986 April and June/July shows variability by several tenths of magnitude
during each observing night and an overall brightening by 0.5 mag from April to
June. On the other hand, the equivalent widths of both He\,{\sc ii} 4686 and
H$\beta$ lines for both data sets are practically the same (C87), and similar
to the mean equivalent widths observed in LMXBs (Smale 1996). However, these
lines can be formed in either the heated hemisphere of the secondary, the
accretion disc, and/or the accretion stream, depending both upon the object
and/or its state of activity, which precludes an unambiguous interpretation of
the spectroscopic periodicity.

Furthermore, we note that C92 do not state whether the 14.8-h periodicity found
by them in the (optically) high-state photometry of Corbet et al.\ (1987) was
present in both 1986 April and June/July or in only one of those data sets. C92
also claim that if the maximum in the light curve (presented in their Fig.\ 6)
corresponds to transit of the compact object in front of its companion, the
radial velocity changes (shown in their Fig.\ 7) are consistent with an origin
near the compact object. However, this is not the case because at both
spectroscopic conjunctions (i.e., the transits of the compact star both in
front and behind the companion) the observed radial velocity should be just the
systemic velocity.

Keeping these doubts in mind let us tentatively assume that the 14.8-h
modulation is still due to the orbital motion. The semi-amplitude, $K_{\rm X} =
78 \pm 13$ km s$^{-1}$ (C92; the uncertainty is 1-$\sigma$) implies the
secondary mass function of
\begin{equation}
f\equiv {M_{\rm c} \sin^3 i\over (1+M_{\rm X}/M_{\rm c})^2}=
0.030^{+0.018}_{-0.012} M_\odot\,,
 \end{equation}
which corresponds to the mass of the primary of
 \begin{equation}
M_{\rm X}= M_{\rm c}\left[ \left(M_{\rm c} /f\right)^{1/2} \sin^{3/2} i
-1\right]\,.
 \end{equation}

From the lack of eclipses, the inclination can be constrained to $i\la 60\degr$
(C87). If we adopt $M_{\rm c}=1M_\odot$ (see above), we obtain $M_{\rm X}\la
5M_\odot$, where the equality corresponds to the minimum $f$ and maximum $i$.
On the other hand, the lack of X-ray dips and fits of models with Compton
reflection (see Section 4) both favour $<60\degr$, and, e.g., $i= 45\degr$
yields $M_{\rm X}\la 3.4M_\odot$. Current theoretical (see, e.g.\ Haensel 1995
for a review) and observational (e.g.\ van Paradijs \& McClintock 1995)
estimates yield the maximum mass of a neutron star of $\la 2M_\odot$. On the
other hand, if we assume that the compact object is a black hole, as strongly
suggested by the similarity of its spectral and timing behaviour to Cyg X-1,
then the presumed $M_{\rm X}\ga 2M_\odot$ corresponds to $M_{\rm c}\ga
0.5M_\odot$ (at the lower limit of $f$). Thus, the mass function of C92
together with the constraints on $M_{\rm c}$ are fully consistent with the
presence of a black hole in GX 339--4 (although the presence of a neutron star
is also allowed). Hereafter, we will adopt fiducial values of $M_{\rm
X}=3M_\odot$ and $i=45\degr$, which are around the middle of the allowed
parameter space.

Then, $P_{\rm orb} = 14.8$ h and $M_{\rm X}+ M_{\rm c} \sim 4 M_{\odot}$ imply
a $1.3 R_{\odot}$ Roche-lobe (tidal) radius for the secondary. A post-main
sequence $1 M_{\odot}$ star with $R \sim 1.3 R_{\odot}$ (and thus filling its
Roche lobe) would have a luminosity of $\sim 1 L_{\odot}$ and an effective
temperature (unheated) of $\sim 5100$ K (Webbink, Rappaport \& Savonije 1983).
Such star at $d=4$ kpc and $\ebv= 1.2$ would have $V \sim 21.6$, $B-V \sim 2$,
$V-R_{\rm c} \sim 1.1$, and $V-I_{\rm c} \sim 2.3$, which correspond to fluxes
below those observed from GX 339--4 (similarly to the case of a $1M_\odot$
main-sequence star, see Fig.\ 2). The expected Roche-lobe overflow rate, $\dot
M$, in such a system is several times $10^{-10} M_{\odot}\, {\rm yr}^{-1}$
(Webbink et al.\ 1983), which is sufficient to power the average X\g\ emission
from GX 339--4 (Rubin et al.\ 1998) of $\sim 2\times 10^{36}$ erg s$^{-1}$
assuming an accretion efficiency of $\eta=0.06$. On the other hand, $\dot M$
can be significantly higher due to X\g\ irradiation of the companion star
(Podsiadlowski 1991; Tavani \& London 1993), which would imply a lower $\eta$.

In this geometry, the secondary star subtends a solid angle of $\sim
0.018\times 4\pi$. Thus, the secondary provides a negligible contribution to
Compton reflection in GX 339--4 (Section 4).

\section{\textbfit{G\lowercase{inga}\/} AND OSSE DATA}
\label{s:data}

GX 339--4 was observed by \ginga\/ on 1991 September 11 and 12 (U94), and by
OSSE 1991 September 5--12 (G95). We have extracted the OSSE data for September
11 and 12 for periods approximately coinciding with the \ginga\/ observations.
The log of the \ginga\/ (from U94) and OSSE observations used here is given in
Table 1. Both the \ginga\/ and OSSE fluxes during September 11--12 were
approximately constant (U94; G95).

\begin{table*} \label{t:log} \centering \caption {The log of observations in
UT. The counts of \ginga\/ and OSSE are for 1.2--29 keV and 50--150 keV energy
ranges, respectively. The OSSE exposure and count rate are normalized to one of
its 4 detectors operating during the observations. }

 \begin{tabular}{lccccccccc}
\hline
&& \multicolumn{4}{c}{\ginga}& \multicolumn{4}{c}{OSSE}\\
Data set &Date& Start & End & Livetime [s] &Counts [s$^{-1}$]
&Start & End & Exposure [s] &Counts [s$^{-1}$]\\
1 & 1991 Sept.\ 11 &$03^{h}35^{m}$ &$05^{h}26^{m}$  &1952 &$1716\pm 4$
&$03^{h}25^{m}$ &$05^{h}47^{m}$ &4795 &$13.14\pm 0.26$\\
2 & 1991 Sept.\ 12 &$00^{h}54^{m}$ &$01^{h}11^{m}$  &384&$1695\pm 5$
&$01^{h}10^{m}$ &$01^{h}59^{m}$ &2488 &$12.71\pm 0.35$\\
\hline
\end{tabular}
\end{table*}

We adopt the current best estimate of the effective area of the Large Area
Counter (LAC) of \ginga\/ (Turner et al.\ 1989)  of 4661 cm$^2$ (D. Smith,
private communication). The usable energy range of \ginga\/ is 1.2--29 keV for
these observations. A 1 per cent systematic error is added in quadrature to the
statistical error in each \ginga\/ channel (as in U94).

The OSSE data are from 50 keV to 1000 keV. They include energy-dependent
systematic errors estimated from the uncertainties in the low-energy
calibration and response of the detectors using both in-orbit and prelaunch
calibration data, and correspond to an uncertainty in the effective area in the
OSSE response. They are most important at the lowest energies ($\sim 3$ per
cent at 50 keV, decreasing to $\sim 0.3$ per cent at $\ga 150$ keV).

\section{ELEMENTARY SPECTRAL MODELS}
\label{s:fits}

In this section, we analyze the \ginga\/ and OSSE data using idealized models
of blackbody emission, and power-law or Comptonization emission from a hot,
isotropic, thermal plasma cloud including their Compton reflection from an
underlying slab. Here, we assume the reflected component is not Comptonized by
the hot plasma. This allows us to characterize spectral components present in
the X\g\ spectrum in a phenomenological, but relatively model-independent, way.
Similar models are widely used in literature to model spectra of other
black-hole sources, e.g.\ Cyg X-1 (Ebisawa et al.\ 1996) or Seyfert 1s (Nandra
\& Pounds 1994), and thus the results of this Section allow a direct comparison
of the obtained parameters with those for other objects. On the other hand, we
introduce more realistic geometries in context of specific models and treat
Comptonization of a fraction of the reflected spectrum as well as energy
balance in Sections \ref{s:geo}--\ref{s:soft} below.

For spectral fits, we use {\sc xspec} (Arnaud 1996) v10. The confidence ranges
of each model parameter are given for a 90 per cent confidence interval, i.e.,
$\Delta \chi^2=2.7$ (e.g.\ Press et al.\ 1992). On the other hand, the plotted
vertical error bars are 1-$\sigma$, the upper limits, 2-$\sigma$, and the
plotted spectral data are rebinned for clarity of the display. Model spectra
are attenuated by $\nh=6\times 10^{21}$ cm$^{-2}$ (Section 2.1). Since the
\ginga\/ exposure is much longer for the data set 1, we discuss below results
obtained with that set. However, we give fit results for both data sets in
Table 2.

As found by U94, the \ginga\/ spectra of GX 339--4 in the hard state consist
of four main components: an underlying power law, a soft excess below $\sim 4$
keV, a continuum due to reflection from the surface of an ionized accretion
disc (e.g.\ Lightman \& White 1988; George \& Fabian 1991), and a fluorescent
Fe K$\alpha$ line.

For modeling Compton reflection, we use inclination-dependent Green's functions
of Magdziarz \& Zdziarski (1995) (instead of the angle-averaged reflection
spectrum used in U94), which assumes an isotropic point source (or,
equivalently, an optically-thin corona) above a slab. The Green's functions are
convolved with an incident continuum (a power law or a thermal Comptonization
spectrum). The reflector inclination is kept at $i=45\degr$ (unless stated
otherwise). We treat the solid angle, $\Omega$, subtended by the reflector as a
free parameter. Values of $\Omega<2\pi$ may correspond either to a truncation
of the reflecting disc, or to substantial Comptonization of the reflected
spectrum (or both). The ionization parameter of the reflector, $\xi=L_{\rm
ion}/nr^2$, is assumed to be uniform in the reflecting material. Here $L_{\rm
ion}$ is defined as the 5 eV--20 keV luminosity in a power law spectrum and $n$
is the density of the reflector located at distance $r$ from the illuminating
source (Done et al.\ 1992). The reflector temperature is kept at $10^6$ K,
which is the highest temperature consistent with the model of ionization
equilibrium used (Done et al.\ 1992). This temperature is consistent with our
estimate of the origin of reflection from a larger area than that giving rise
to the observed soft excess, see Section 5.4. The abundances of both the
reflector and the interstellar medium are from Anders \& Ebihara (1982) except
that the relative Fe abundance in the reflector, $\af$, is a free parameter.
The ion edge energies and opacities are from Reilman \& Manson (1979, used in
U94) except that now the Fe K-edge energies are from Kaastra \& Mewe (1993).
The continuum reflection is accompanied by an Fe K$\alpha$ line, which we model
as a Gaussian centered at an energy, $\efe$, with a width, $\sfe$, and a flux,
$\ife$.

\begin{table*} \centering \caption {The parameters of the fits in Section 4 to
the \ginga\/ and OSSE data. For each data set, the consecutive fits are to the
4.1--29 keV, 4.1--1000 keV, and 1.2--1000 keV ranges, respectively. $L$ and
$L_{\rm bb}$ are the unabsorbed bolometric luminosities of the total spectrum
and of the blackbody component only, respectively, in units of $10^{37}$ erg
s$^{-1}$. $A$ is the 1-keV normalization of the primary continuum (in
cm$^{-2}$\,s$^{-1}$\,keV$^{-1}$), $\xi$ is in erg cm s$^{-1}$, $\ife$ is in
$10^{-3}$ cm$^{-2}$ s$^{-1}$, $kT$, $\tbb$ are in keV, and $\wfe$ is in eV. }

 \begin{tabular}{lccccccccccccc}
\hline

No. & $A$ & $\Gamma$ &$kT$ &$L$ &$\Omega/2\pi$ &$\xi$ &$\af$ &$\tbb$
&$L_{\rm bb}$ &$\efe$ &$\ife$ &$\wfe$ &$\chi^2$/dof \\

\multicolumn{14}{c}{1991 September 11}\\
1a & 0.75 & $1.75^{+0.02}_{-0.03}$ & -- & -- & $0.37^{+0.06}_{-0.05}$ &
$120^{+160}_{-70}$ & $2.5^{+1.2}_{-0.8}$ & -- & -- & $6.51_{-0.32}^{+0.31}$ &
$1.5_{-0.8}^{+0.7}$ & $49^{+26}_{-26}$ & 11.7/27 \\
1b & 0.75 & $1.76^{+0.02}_{-0.01}$ &$56^{+6}_{-6}$ & --
  &$0.43^{+0.03}_{-0.07}$ &$110^{+120}_{-50}$ &$2.9^{+0.1}_{-0.7}$
  & -- & --   & $6.51^{+0.16}_{-0.35}$ &$1.3^{+0.8}_{-0.7}$ & $45^{+26}_{-25}$
  &45.2/79\\
1c & 0.60 & $1.77^{+0.01}_{-0.01}$ &$57^{+7}_{-5}$ &3.12
  &$0.44^{+0.06}_{-0.06}$ &$90^{+90}_{-40}$ &$3.0^{+0}_{-0.7}$
  &$0.25^{+0.02}_{-0.03}$ &0.29
  &$6.54^{+0.33}_{-0.36}$ &$1.3^{+0.7}_{-0.8}$ &$44^{+25}_{-26}$ &47.8/82\\
\multicolumn{14}{c}{1991 September 12}\\
2a & 0.73 & $1.74^{+0.03}_{-0.04}$ & -- & -- & $0.25^{+0.08}_{-0.08}$ &
$230^{+510}_{-180}$ & $1.6^{+2.2}_{-0.8}$ & -- & -- & $6.56_{-0.37}^{+0.34}$ &
$1.4_{-0.9}^{+0.9}$ & $48^{+31}_{-32}$ & 22.4/27 \\
2b & 0.75 & $1.76^{+0.02}_{-0.02}$ & $53^{+8}_{-7}$ & --
  &$0.29^{+0.09}_{-0.08}$ &$170^{+310}_{-130}$ &$2.0^{+1.0}_{-0}$
  & -- & --   & $6.57^{+0.37}_{-0.41}$ &$1.3^{+0.9}_{-0.9}$ & $44^{+31}_{-31}$
&61.5/79\\
2c & 0.55 & $1.76^{+0.02}_{-0.03}$ &$52^{+7}_{-6}$ &2.93
  &$0.29^{+0.08}_{-0.08}$ &$170^{+300}_{-120}$ &$2.0^{+1.0}_{-0}$
  &$0.27^{+0.03}_{-0.03}$ &0.21
  &$6.56^{+0.35}_{-0.38}$ &$1.4^{+0.8}_{-1.0}$ &$46^{+30}_{-31}$ &62.5/82\\
\hline
\end{tabular}
\end{table*}

We first fit the data set 1 from \ginga\/ in the 4.1--29 keV range only, where
a contribution from the soft excess was found to be negligible by U94. We first
use a model consisting of a power law and a line. This model provides a very
poor description of the data, with $\chi^2=209/30$ d.o.f. The pattern of
residuals is characteristic of the presence of Compton reflection, as shown in
Fig.\ 2a of U94. Indeed, adding a reflection component improves the fit
dramatically, reducing $\chi^2$ to $11.7/27$ d.o.f., see fit 1a in Table 2
(which also give corresponding results for the data set 2) and Fig.\ 3. Compton
reflection is present at very high statistical significance, with a probability
of $<10^{-16}$ that adding this component were not required by the data (as
obtained using the F-test). The statistical significances of the reflector
being ionized, overabudant in Fe, and an Fe K$\alpha$ being present in the
spectrum correspond to the probability that the resulting fit improvement
were by chance of $<10^{-4}$, $3\times 10^{-7}$ and $3\times 10^{-4}$,
respectively. The width of the line is only weakly constrained, $\sfe\la 0.5$
keV at 1$\sigma$, and thus we keep it fixed at 0.1 keV hereafter.

We comment here on the low reduced $\chi^2$ obtained for the data set 1. This
is due to the 1 per cent systematic error added to the statistical error. This
value appears to be an overestimate of the residual inaccuracy of the
calibration of the \ginga\/ LAC, for which 0.5 per cent appears to be a better
estimate. However, we have retained here the statistical error adopted by U94
in order to enable direct comparison with their results. This leads to
uncertainties on the parameters of our fits being more conservative than
necessary, but it affects neither the best-fit values nor the conlusions
regarding the spectral components present in the \ginga\/ spectrum. E.g.\
Compton reflection is still found at a very high significance, see above. The
addition of the large systematic error also leads, in some cases, to a
relatively low reduced $\chi^2$ for models showing systematic (rather than
random) departures from the \ginga\/ data, which cases we discuss below
individually. On the other hand, the reduced $\chi^2$ is significantly larger
for the data set 2, see Table 2. This is due to its much shorter exposure,
resulting in typical statistical errors $>1$ per cent, which then reduces the
relative importance of the systematic errors.

We have then tested whether there is any indication in the data for the
presence of kinematic and relativistic smearing of the Fe K$\alpha$ and the
reflection continuum. Such smearing would be present if those spectral
components were formed close to the central black hole, which effect has been
found in \ginga\/ observations of the X-ray novae GS 2023+338 and Nova Muscae
(\.Zycki, Done \& Smith 1997, 1998). We have used a model in which the line and
the static reflection continuum were convolved with the disc line profile of
Fabian et al.\ (1989), similar to the model of \.Zycki et al.\ (1997, 1978).
However, we have found no fit improvement ($\Delta\chi^2=-0.6$) allowing for
the smearing, although the data also do not allow us to rule it out.

The relative normalization of the reflected components, $\Omega/2\pi$,
corresponds to $\sim 0.6$ of those obtained by U94. This is explained mostly by
their use of the angle-averaged spectrum, which underestimates the actual
reflection spectrum for angles $<65\degr$, which effect increases at energies
$\ga 15$ keV (Magdziarz \& Zdziarski 1995). Accordingly, the fit to the data
set 1 with the angle-averaged reflection in U94 shows strong positive residuals
above $\sim 20$ keV (Fig.\ 2b in U94), which systematic residuals disappear
completely in the present fit, see Fig.\ 3. Also, we obtain $\xi$ about 2--3
times less than those of U94. However, the ionization state depends on both
$\xi$ and the reflector temperature, and the difference is due to the
assumption of U94 that the latter is $10^5$ K (which seems unlikely in inner
regions of luminous compact objects in binaries).

We then investigate the issue of the Fe abundance. We have fitted the earlier
four hard-state observations of GX 339--4 by \ginga\/ in 1989--90 (U94), and we
found all of them consistent with $2\leq \af\leq 3$, which is also the case for
the present data. Therefore, we constrain $\af$ to this range hereafter.

In agreement with U94, we find a strong excess below 4 keV, with the fluxes in
the 1.2--1.7 keV and 1.7--2.3 keV bands about 40 and 20 per cent, respectively,
higher than the extrapolation of the model fitted to the 4.1--29 keV range, as
shown in Fig.\ 3. Although the exact form of the excess depends on the adopted
$\nh$,  we find it to be significant, with the respective 1.2--1.7 keV and
1.7--2.3 keV relative excess of 25 and 12 per cent even for $\nh=4\times
10^{21}$ cm$^{-2}$, which is much below the optical and X-ray estimates
(Section 2.1). Thus, we conclude that the presence of the soft X-ray excess is
not an artefact of our choice of $\nh$.

\begin{figure} \begin{center} \leavevmode \epsfxsize=7cm
\epsfbox{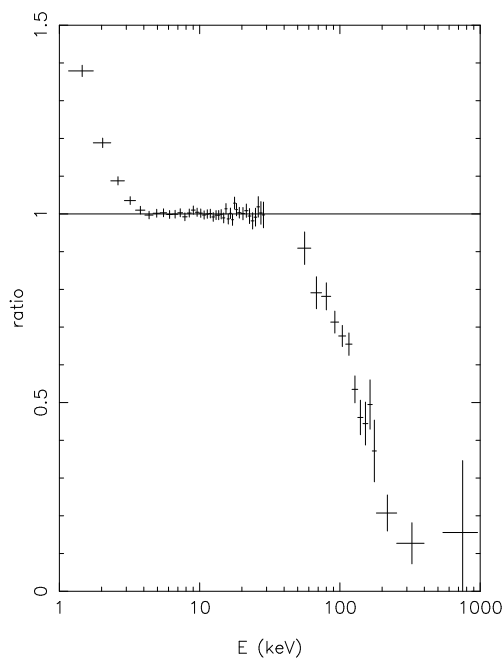} \end{center} \caption{The ratio of the data of 1991
September 11 from \ginga\/ and OSSE to the model consisting of a power law,
Compton reflection, and an Fe K$\alpha$ line fitted to the \ginga\/ data in the
4.1-29 keV energy range. } \end{figure}

At 50 keV, the extrapolated \ginga\/ fit predict the flux about 10 per cent
higher than that observed by OSSE. Above 50 keV, the OSSE data show a strong
cutoff, as illustrated in Fig.\ 3. We test whether the cutoff can be modelled
by thermal Comptonization, similarly to the case of Cyg X-1 (Gierli\'nski et
al.\ 1997, hereafter G97) and Seyfert AGNs (e.g.\ Zdziarski et al.\ 1997,
hereafter Z97). We fit first the \ginga\/ data above 4.1 keV jointly with the
OSSE data. The incident continuum is fitted by thermal Comptonization in a
spherical cloud with a central source of soft seed photons having a blackbody
distribution, see Zdziarski, Johnson \& Magdziarz (1996). As discussed in that
paper, the solution of the Kompaneets equation with relativistic corrections
used there leads to an overestimation of the actual plasma temperature, $T$,
when the Thomson depth of the plasma, $\tau$, is $\la 2$. Therefore, we correct
here the values of $T$ obtained from the Kompaneets equation by a function (R.
Misra, private communication) obtained by comparison with corresponding Monte
Carlo results (Zdziarski et al.\ 1996). The second parameter of the model is
the asymptotic power-law index in X-rays, $\Gamma$, which is related to the
(geometry-dependent) $\tau$ by \begin{equation} \tau\simeq \Theta^{-1/2} \left[
\left(\Gamma+{1\over 2}\right)^2 - {9\over 4}\right]^{-1/2}, \end{equation}
where $\Theta\equiv kT/m_{\rm e} c^2$ and $m_{\rm e}$ is the electron mass. In
fits, we initially assume the seed photon temperature of $\tbb =0.1$ keV; as
long as $\tbb$ is much less than the minimum energy in the fitted data, its
choice does not affect the fit results. This model provides a very good
description of the data (see fits 1b and 2b in Table 2). The best fit
corresponds to $\tau\simeq 1.8$.

As an independent check of our results, we have also used a model of Coppi
(1992), which treats thermal Comptonization using the formalism of escape
probability. That model yields $kT\simeq 48$ keV, $\tau\simeq 1.93$,
$\Omega/2\pi=0.42$ as the best fit ($\chi^2=49/79$ d.o.f.), rather similar to
the fit 1b in Table 2. On the other hand, we find that the high-energy cutoff
seen it the data is poorly modelled by an e-folded power law, which, apart from
its mathematical simplicity, does not correspond to any physical model. We find
that model fits the data much worse than thermal Comptonization, with $\Delta
\chi^2=+19$ resulting in a systematic pattern of residuals in the OSSE data.
This argues for thermal Comptonization being indeed the process giving rise to
the intrinsic spectrum.

\begin{figure}
\begin{center}
\leavevmode
\epsfxsize=8.4cm \epsfbox{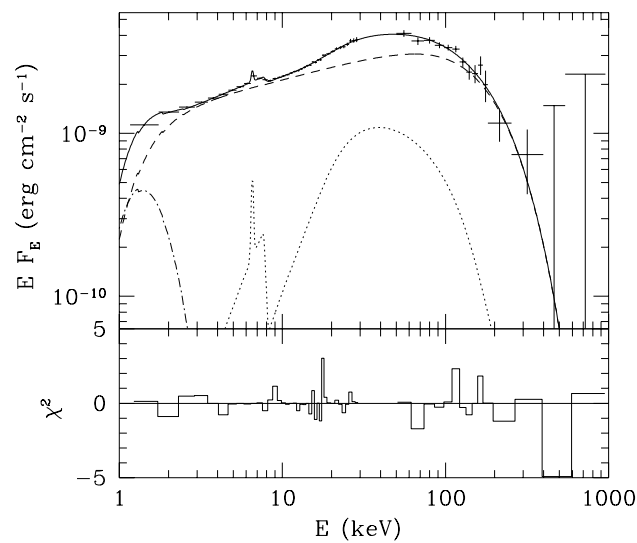}
\end{center}
\caption{The X\g\ spectrum (crosses) of GX 339--4 observed simultaneously by
\ginga\/ and OSSE on 1991 September 11 (the spectrum of September 12 is very
similar and thus it is not shown). The data are fitted by a model consisting
of blackbody radiation (dot-dashed curve) providing seeds for thermal
Comptonization in a hot plasma (dashed curve). The hard radiation of the hot
plasma is Compton-reflected from some cold medium, which component is
shown by the dotted curve. The solid curve gives the sum. Note that the
observed spectrum is attenuated at soft X-rays in the interstellar medium. The
bottom panel gives the contribution to the total $\chi^2$ from separate data
bins multiplied by the sign of (data $-$ model).
 }
\end{figure}

We have considered the effect of possible inaccuracy of the relative
normalization of the effective area of \ginga\/ and OSSE. We find, however, no
fit improvement, $\Delta \chi^2=-0.1$, when the relative normalization of OSSE
with respect to \ginga\/ is allowed to be free (reaching 0.98 for both data
sets). Therefore, we keep it at unity hereafter.

We then check the effect of changing the disc inclination. We find that
allowing a free $i$ leads to a negligible fit improvement: $\chi^2= 44.3/78$
d.o.f.\ at $i=23_{-23}^{+38}$ deg. Thus, we keep it at $i=45\degr$ (see Section
2.3) hereafter. At the largest $i$ allowed by the fit, $\Omega/2\pi =0.62$ and
$\wfe=35$ eV.

We then consider the soft X-ray excess. In the hard state of Cyg X-1, the soft
excess contains a blackbody component with $\tbb\approx 0.14$ keV (e.g.\
Ebisawa et al.\ 1996). In addition, the \asca\/ data for Cyg X-1 show also a
break around 3--4 keV with the power law below the break being softer by
$\Delta \Gamma \sim 0.4$ on average (Ebisawa et al.\ 1996). The physical origin
of that spectral break is unclear. We find that the soft X-ray excess in our
data is much better fitted by an additional blackbody component than by a
broken power law ($\Delta \chi^2=5$). (Our data extend down to 1.2 keV only
with a low energy resolution, and thus do not allow us to determine the
presence of more than one spectral component of the soft excess.) In the model
with a blackbody component, we set its temperature equal to the blackbody
temperature of the seed photons in the Comptonization source. (We postpone
examining the energy balance of the source to Sections 5.1--5.2 below.) This
model gives an excellent description of the \ginga/OSSE data in the entire
energy range, 1.2--1000 keV, see Fig.\ 4 and fits 1c and 2c in Table 2. Still,
according to the above discussion, the fitted value of $\tbb$ is relatively
uncertain; if there were a low-energy break in the power-law spectral component
(as in Cyg X-1), $\tbb$ would be lower.

\section{PHYSICAL IMPLICATIONS}

\subsection{Geometry and energy balance}
\label{s:geo}

We now consider thermal Comptonization, Compton reflection and reprocessing in
realistic geometries approximating the structure of an accretion flow. We take
into account anisotropy of both seed and Compton-scattered photons,
Comptonization of the reflected photons that return to the hot plasma, and
energy balance.

In general, thermal Comptonization takes place in a hot plasma cloud, and the
resulting spectrum is Compton-reflected from a cold medium. A major part of the
Comptonized flux incident on the cold medium is bound-free absorbed rather than
reflected, which gives rise a flux of reprocessed soft photons with a spectrum
close to a blackbody. In turn, a geometry-dependent fraction of the reflected
and the blackbody emissions returns to the hot plasma. These blackbody photons
then serve as seeds for thermal-Compton upscattering (in addition to any other
soft photons present). Those reflected photons that return to the hot plasma
are upscattered, which leads to a strong suppression of that component for
$\tau\ga 1$.

A geometry can be ruled out either if the resulting model spectrum does not
provide a good fit to the observed spectrum, or if energy balance between the
hot plasma and the source of seed photons cannot be achieved. The condition of
energy balance can be expressed as, e.g.\ the power in soft photons incident on
the hot plasma (hereafter seed photons) times the Comptonization amplification
factor (a function of $kT$ and $\tau$) being equal the power emitted by the hot
plasma, $L_{\rm hot}$. The flux in seed photons arises due to both internal
dissipation and reprocessing of the Comptonized radiation. Then, a sufficient
condition to reject a model is (i) the Comptonized flux (at fitted $kT$ and
$\tau$) larger than that observed, due to a strong flux in seed photons arising
from reprocessing. The condition (i) is equivalent to either statement: (ii) at
the fitted $\tau$ and the observed hard flux, Compton cooling by the seed
photons would result in a $kT$ less than that obtained from fitting, or (iii)
at the fitted $kT$ and the observed hard flux, energy balance can be achieved
for values of $\tau$ less than that fitted to the data (e.g.\ Haardt \&
Maraschi 1993; Stern et al.\ 1995). Below, we first obtain $kT$ and $\tau$ by
spectral fitting, and then a posteriori check the energy balance by finding the
equilibrium temperature corresponding to the fitted $\tau$ in given geometry
[condition (ii) above].

In this section, we neglect spectral constraints from the soft X-ray excess
(but return to this issue in Section \ref{s:soft} below) because the origin of
the observed soft excess is relatively uncertain, see the end of Section 4.
Therefore, we consider now only the data above 4 keV, and fix the seed-photon
temperature at $\tbb=0.25$ keV (as obtained in Table 2), which is about the
highest seed-photon temperature allowed by the data regardless of the actual
form of the soft excess. For a lower $\tbb$ at a given seed-photon flux from
reprocessing, the predicted Comptonized flux would increase, which would in
turn lead to rejection of a larger class of models [see condition (i) above].
Thus, setting $\tbb= 0.25$ keV is the most conservative assumption for model
rejection.

To model Comptonization in anisotropic geometries, we use the iterative
scattering method of Poutanen \& Svensson (1996). For Compton reflection, we
use the same method as in Section 4 (Magdziarz \& Zdziarski 1995). Fig.\ 5
shows the geometries considered in Sections \ref{s:geo}--\ref{s:soft}.

\begin{figure} \begin{center} \leavevmode \epsfxsize=5.2cm
\epsfbox{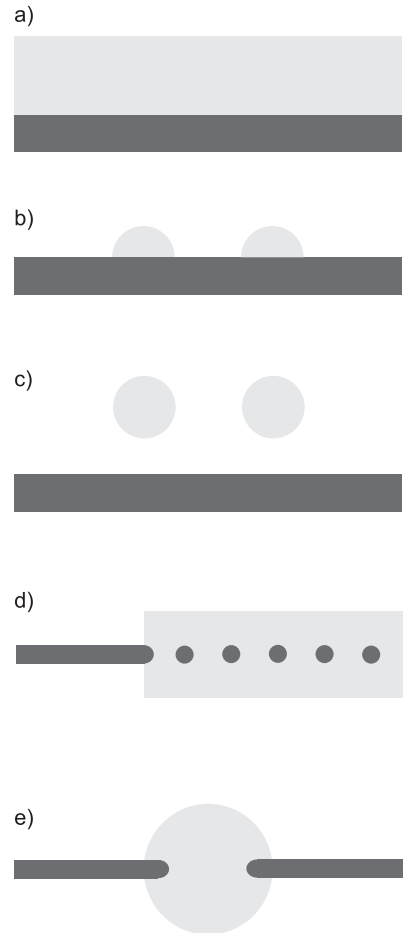} \end{center} \caption{Geometries of the X\g\ source
considered in Sections 5.1--5.2. Dark and light shades denote cold medium and
hot plasma, respectively. {\it (a)\/} A homogeneous corona above a disc, {\it
(b)\/} active regions located on the disc, {\it (c)\/} active regions detached
from the disc, {\it (d)\/} cold clouds in the midplane of a hot disc,
and {\it (e)\/} a hot central sphere surrounded by a cold
disc, which may penetrate some distance within the sphere. Our
preferred models are {\it d, e}. } \end{figure}

We first consider a homogeneous corona (with the vertical optical depth of
$\tau$) covering a cold slab (Haardt \& Maraschi 1993), see Fig.\ 5a. In this
model, all Compton-reflected radiation emitted by the cold slab is Comptonized
in the hot corona, which leads to a suppression of this component to a level
much less than that present in the data. Still, the model can provide a good
spectral fit if we allow an additional Compton-reflection component
(with $\Omega/2\pi\sim 0.3$) due to an outside cold medium, e.g.\ an outer cold
disc. The fitted plasma parameters are then $kT= 55^{+8}_{-7}$ keV and $\tau=
1.2^{+0.2}_{-0.3}$ ($\chi^2= 45/79$ d.o.f.). However, this model is strongly
ruled out by the requirement of energy balance. Namely, the reprocessed flux
from the slab is so strong that it would cool down the corona to $\sim 27$ keV.
Allowing for internal dissipation in the cold slab (presumably a cold accretion
disc) would worsen the discrepancy.

We then consider a patchy corona geometry (Galeev, Rosner \& Vaiana 1979;
Haardt, Maraschi \& Ghisellini 1994; Stern et al.\ 1995). The hot plasma is
postulated here to form active regions above a cold accretion disc. First, we
assume the active regions to form hemispheres located directly on the surface
of the cold disc, see Fig.\ 5b. We obtain a good spectral fit ($\chi^2=47/79$
d.o.f.) for $kT=64^{+9}_{-8}$ keV and the radial $\tau= 2.1^{+0.2}_{-0.3}$.
However, similarly to the case of a homogeneous corona, we find cooling by the
seed photons arising from reprocessing in an underlying part of disc is so
large that the plasma with the fitted $\tau$ would cool to $kT=45$ keV, i.e.,
it cannot sustain the temperature implied by the data in this model.

The cooling, however, can be reduced if the active regions are at some distance
above the disc, as shown in Fig.\ 5c. We find that at the height of about one
radius, there is an energy balance between the disc without internal
dissipation and the hot region. The height has to be larger if there is
internal dissipation in the disc. However, this model implies much more Compton
reflection (from disc regions surrounding the active region) than observed.
Even if we assume that the underlying disc is truncated, which would reduce the
amount of reflection, the spectral fit of this model is much poorer than other
models, $\chi^2=53/79$ d.o.f.\ (for $kT=62^{+14}_{-4}$ keV, $\tau=
2.2^{+0.2}_{-0.3}$, $\Omega/2\pi= 0.71^{+0.10}_{-0.07}$). The cause of the bad
fit is the presence of an anisotropy break in the model spectrum, i.e., a
defficiency of low-energy photons emitted upward due to an anisotropic
suppression of the first order of Compton scattering (Haardt \& Maraschi 1993;
Poutanen \& Svensson 1996). This effect is not seen in the data, and thus this
model yields systematic departures from the \ginga\/ data at low energies. On
the other hand, a good fit can be obtained if $kT_{\rm bb} \la 0.1$ keV is
assumed (which yields $kT=57^{+6}_{-4}$ keV, $\tau= 2.5^{+0.3}_{-0.2}$,
$\Omega/2\pi= 0.67^{+0.07}_{-0.09}$, $\chi^2=45/79$ d.o.f.), in which case the
effect of the anisotropy break on the model spectrum does not extend above 4
keV. The weak Compton reflection required by the model could then correspond to
the cold disc being truncated around the active region, and the origin of the
soft X-ray excess has to be due to an effect different than the disc blackbody
emission. Thus, in this respect, the model is in principle possible.

However, U94 found that $\Omega$ in GX 339--4 correlates positively with
$\Gamma$, which behaviour is opposite to that expected in the patchy corona
model.  We first note that since $\tau\ga 2$ in this model, virtually all
Compton-reflected photons passing through an active region are removed from the
observed reflected spectrum regardless of the exact value of $\tau$. On the
other hand, a decrease of the observed $\Omega$ can occur when the height of
the active region decreases, due to fewer reflected photons being able to
escape without hitting the hot plasma. This also will soften the spectrum due
to the increased cooling by more blackbody photons emitted by the disc
intercepted by the hot plasma. The observed opposite correlation provides then
strong evidence against the patchy corona model.

On the other hand, the data can be well modelled in all respects by a geometry
with cold clouds inside a hot slab (S76; Celotti, Fabian \& Rees 1992; Kuncic,
Celotti \& Rees 1997; Collin-Souffrin et al.\ 1996; Krolik 1998) surrounded by
a cold disc, see Fig.\ 5d.
We assume that the clouds are located in the slab midplane and cover a fraction
of $f_{\rm c}$ of the midplane. If there is neither dissipation in the cold
clouds nor outside seed photons, a solution satisfying both the energy balance
and spectral constraints corresponds to $f_{\rm c}\simeq 0.3$, see Fig.\ 6. The
plasma parameters are then $kT=51^{+7}_{-3}$ keV and $\tau
=0.95^{+0.13}_{-0.15}$ corresponding to the half-thickness of the slab (at
$\chi^2=45/79$ d.o.f.). Compton reflection by the cold clouds is attenuated by
the hot plasma and thus we need the presence of additional Compton reflection
from an outside matter with $\Omega/2\pi \simeq 0.4$ (see Table 2). This can
occur due to reflection of the radiation of the hot flow by an outside cold
disc. The covering factor by the cold clouds will be $f_{\rm c}<0.3$ if
additional soft photons from the outside cold disc and/or from dissipation in
the cold clouds (with the luminosity $L_{\rm bb, intr}$) enter the hot flow,
see Fig.\ 6.

As argued by S76, the clouds can be formed by the radiation-pressure induced
instability of an optically-thick disc (Lightman \& Eardley 1974; Shakura \&
Sunyaev 976). The clouds are in pressure equilibrium with the hot medium.

\begin{figure*} \begin{center} \leavevmode \epsfxsize=10.4cm
\epsfbox{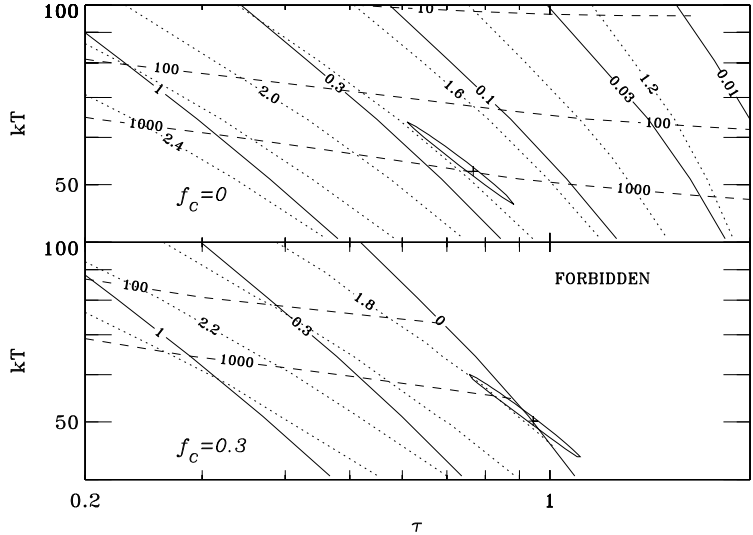} \end{center} \caption{Relations between the electron
temperature, $kT$, and the half-thickness optical depth, $\tau$, of a hot slab,
see Fig.\ 5d and Section 5.1. The covering factor of cold clouds in the
mid-plane of the hot slab  is $f_{\rm c}\rightarrow 0$ (upper panel) and
$f_{\rm c}=0.3$ (lower panel). The solid curves correspond to constant $L_{\rm
bb,intr}/L_{\rm hot}$, i.e., the ratio of the internally generated and/or
external soft luminosity to the dissipation rate in the hot disc. These
relations do not depend on the amount of $e^{\pm}$ pairs. The dashed curves
correspond to a constant local hard compactness, $\ell$ (assuming
pair-dominated thermal plasma, see Section 5.6). The dotted curves correspond
to a constant intrinsic (i.e., without Compton reflection) spectral index
$\Gamma$ in 4--20 keV range. Crosses are the best fits to the data and
elongated confidence contours are for 90 per cent confidence for 2 parameters
($\Delta\chi^2=4.61$). For $f_{\rm c}\rightarrow 0$, the seed-photon luminosity
entering the hot slab is $\sim 0.23$ of the heating rate, $L_{\rm hot}$. The
covering factor of $f_{\rm c}=0.3$ is the largest energetically possible
consistent with the data. In that case, no internal dissipation in the cold
clouds and no external soft photons are allowed, and cooling is provided by
reprocessed radiation only. } \end{figure*}

We have also tested a similar geometry with the hot plasma forming a central
hot sphere (rather than a slab) surrounded by a cold disc (e.g.\ Poutanen,
Krolik \& Ryde 1997), see Fig.\ 5e. If the cold disc does not penetrate into
the hot cloud, $kT=57_{-3}^{+6}$ keV and the radial $\tau =2.0_{-0.2}^{+0.1}$
($\chi^2=47/79$ d.o.f.), and internal dissipation in the cold disc is required
to provide enough seed soft photons. The model also predicts less Compton
reflection than observed, which problem may be solved by flaring of the outside
disc. The cold-disc solution of Shakura \& Sunyaev (1973) does, in fact,
correspond to a flared disc, and the illumination by the central hot source
will further increase the amount of flaring. We note that a central hot sphere
is a poor approximation to the actual geometry of a hot accretion flow with
substantial cooling (see Section \ref{s:accretion}), in which the scaleheight
is maximized at the outer boundary rather than at the center, which effect will
also lead to an increase of the amount of reflected X-rays. We also find that
models with the cold disc penetrating into the hot cloud down to $\ga 0.7$ of
the sphere radius can also fit the data and satisfy the energy balance, but
with less or no dissipation in the cold disc. An overlap between the cold and
hot phases increases the amount of Compton reflection in the model and thus
reduces the $\Omega/2\pi$ needed from flaring of the outside cold disc.

The $\Omega$ vs.\ $\Gamma$ correlation found by U94 may be explained naturally
by geometries shown in Figs.\ 5d, e. In these geometries, an increase of the
area of the outside soft X-ray emitter leads to an increase of the seed-photon
flux incident on the hot plasma, which in turn leads to more cooling and
softening of the X-ray spectrum, i.e., an increase of $\Gamma$. The same effect
leads to more Comton reflection, i.e., an increase of $\Omega$.

Summarizing this section, our best models consist of a central hot region
surrounded by a cold disc (Figs.\ 5d, e). These models require the presence of
an additional Compton-reflection component, e.g.\ from an outer cold disc.
Among those 2 models, a hot disc with coldclouds insidesurrounded by a cold
disc (Fig.\ 5d) is more likely to be formed by an accretion flow with
substantial cooling (see Section \ref{s:accretion} below). Both models explain
the $\Omega$-$\Gamma$ correlation. On the other hand, a model consisting of
active regions at some height above a cold disc (Fig.\ 5c) is marginally
possible, but it requires the temperature of blackbody photons lower than that
inferred from the soft X-ray excess and a truncation of the underlying disc,
and it predicts an $\Omega$-$\Gamma$ correlation opposite to that observed.
Models with a homogeneous corona and with the active regions on the disc
surface (Figs.\ 5a, b) are strongly ruled out by the requirement of energy
balance.

\subsection{The origin of the soft X-ray excess}
\label{s:soft}

In this Section, we fit thedata in the full 1.2--1000 keV range, i.e.,
including the range $<4$ keV showing the soft X-ray excess. Based on the
results of Section \ref{s:geo}, the most likely geometry of the hot plasma is a
hot inner disc mixed with cold clouds.  We first check whether the soft excess
can be accounted for just by the emission of the cold clouds,which we find
energetically viable. However, this model fits the data much worse than our
baseline model of Table 2, $\Delta\chi^2=+9$, even for the maximum possible
covering factor of the cold clouds (see Fig.\ 6), $f_{\rm c}=0.3$.

On the other hand, the hot disc is surrounded by a cold outer disc beyond a
transition radius,$R_{\rm tr}$ (Fig.\ 5d), which contributes to the soft
excess. We have thus added to our spectral model an integral of blackbody
spectra over the disc surface at $R\geq R_{\rm tr}$. The local color
temperature is in general larger than the effective one, and  the ratio of the
two for best estimates of the system parameters (Section 2) is $T_{\rm
color}/T_{\rm eff} \simeq 1.8$ (Shimura \& Takahara 1995). Following Makishima
et al.\ (1986), we neglect here the boundary-condition factor $J(r)\equiv
1-(6/r)^{1/2}$, where $r\equiv Rc^2/G M_{\rm X}$, which is a good approximation
for a large enough $r_{\rm tr}$. We find that due to the limited energy range
of the \ginga\/ data, our fits cannot determine the value of $R_{\rm tr}$. On
the other hand, $R_{\rm tr}$ can be obtained from the area covered by the cold
clouds emitting blackbody spectrum with the fitted color temperature, $\tbb$.
This yields $R_{\rm tr} \simeq 6\times 10^7$ cm at the best fit. The main model
parameters are given in Table 3, where $\tin$ is the cold disc temperature at
$R_{\rm tr}$, and Compton reflection is from the cold disc. The model spectrum
is shown in Fig.\ 7. The bolometric luminosity is $4\times 10^{37}$ erg
s$^{-1}$, and the ratio of the blackbody luminosity of the cold disc to the
total one (but without including the outside reflection component) is $\simeq
0.30$.

\begin{table*} \centering \caption {The main parameters of of our best model,
consisting of a hot inner disc with internal cold clouds surrounded by a cold
disc, fitted to the data set 1. See Figs.\ 5d, 7 and Section 5.2. Temperatures
are in units of keV. }

 \begin{tabular}{lccccc}
\hline

$\tau$ &$kT$ &$f_{\rm c}$ &$\tbb$ & $kT_{\rm tr}$ & $\chi^2/$dof \\

$0.88^{+0.10}_{-0.10}$ & $52^{+1}_{-3}$ & $0.26^{+0.04}_{-0.11}$ &
$0.33^{+0.02}_{-0.04}$ &$0.21^{+0.01}_{-0.01}$ & 48.6/81 \\

\hline \end{tabular} \end{table*}

\begin{figure} \begin{center} \leavevmode \epsfxsize=8.4cm
\epsfbox{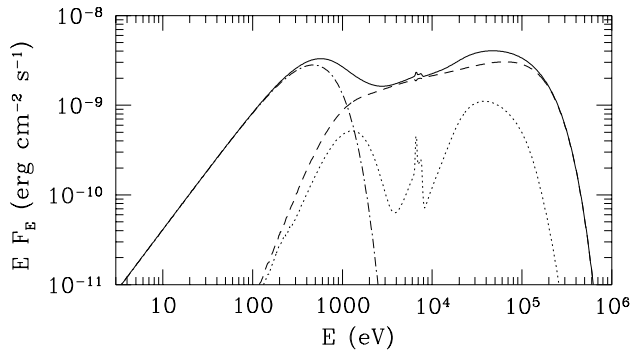} \end{center} \caption{The unabsorbed model spectrum
corresponding to the fit to the data of our best model, a hot disc surrounded
by a cold disc, see Fig.\ 5d. The dot-dashed curve corresponds to the
disc-blackbody emission, the dashed curve gives the Comptonized emission in the
hot disc, and the dotted curve gives the sum of the blackbody emission of the
cold clouds within the hot disc and the reflected component from both the
clouds and the cold disc. The solid curve gives the total spectrum. }
\end{figure}

We point out that we have neglected the effect of heating by the central source
on the radial temperature dependence of the outside cold disc. This may
significantly soften the power-law part of the disc-blackbody spectrum shown in
Fig.\ 7 (e.g.\ Vrtilek et al.\ 1990).

A model alternative to the hot inner disc that was found by us marginally
possible (Section \ref{s:geo}) is active regions at some height above the disc
(Fig.\ 5c). In this model, the soft excess can be accounted for by the
blackbody emission of the disc region underneath the active region. Due to the
intensive heating by the X-rays, the local temperature can be much above the
temperature of a disc with internal dissipation only. From the strength of the
blackbody component (Table 2), the characteristic size of the heated region of
the disc is $\sim 10^7 (T_{\rm color}/T_{\rm eff})^2$ cm. However, when we
approximate the emission of the underlying region of the disc as a blackbody, a
very poor fit to the data is obtained, $\chi^2=93/82$ d.o.f. This provides one
more argument against this model.

Summarizing this section, the most likely origin of the soft excess we have
found is from an outer cold accretion disc with an additional contribution from
cold clouds within the hot disc, see Figs.\ 5d and 7.

\subsection{Accretion disc solutions}
\label{s:accretion}

The best overall model found in Sections \ref{s:geo}--\ref{s:soft} consists of
a hot plasma slab with the half-thickness corresponding to $\tau\la 1$, and
$kT\ga 50$ keV. Physically, this geometrical model likely corresponds to a hot
accretion disc. We note that indeed the best-fit parameters of the hot slab are
close to those predicted by the two-temperature, hot disc model of S76. In that
model, the local gravitational energy is converted into the thermal energy of
ions (at a temperature $T_{\rm i}\gg T$), which is then transferred to
electrons by Coulomb interactions. The electrons, in turn, radiate away their
energy by Compton upscattering seed photons irradiating the plasma. As found by
e.g.\ Ichimaru (1977), Abramowicz et al.\ (1995) and Narayan \& Yi (1995),
advection of hot ions to the black hole leads to another branch of the hot
solution as well as it limits the accretion rate possible in the hot flow.

Detailed properties of the solution in the vicinity of the maximum accretion
rate are studied by Zdziarski (1998, hereafter Z98). That study follows S76 in
characterizing the flow by a value of the Compton parameter, $y\equiv 4\Theta
\max(\tau, \tau^2)$ (e.g.\ Rybicki \& Lightman 1979), which approximately
determines the X-ray spectral index, $\Gamma$. This is equivalent to assuming
that flux in the seed photons is such that it gives rise to an X-ray spectrum
with that $\Gamma$.

Other parameters of the flow are $\dot M$ and the viscosity parameter, $\alpha$
(Shakura \& Sunyaev 1973). [The rate of advection is characterized by a
parameter $\xi_{\rm adv}\simeq 1$, e.g.\ Chen, Abramowicz \& Lasota (1997).]
Z98 has obtained that the optical depth (corresponding to the scaleheight, $H$)
of the flow weakly depends on $R$, and that $\tau$ at the local maximum rate,
$\mmax$, and the maximum $\tau$ (see Fig.\ 8) are,
\begin{equation} \label{eq:tau_max}
\tau(\mmax)\simeq 1.22 y^{3/5} \alpha^{2/5}, \quad \tau_{\rm max}\simeq 1.54
y^{3/5} \alpha^{2/5}, \end{equation}
respectively. Those values are reached at $r\simeq 13$, at which the rate of
dissipation per logarithmic radius interval is also close to maximum. At
$\mmax$, advection carries about half the dissipated power and $H/R \sim 1$.

The luminosity of GX 339--4 during the observations reported is about the
maximum observed in the hard state (Harmon et al.\ 1994; Rubin et al.\ 1998).
Thus, we can relate the above $\tau$ with that obtained from fitting the slab
model in Section \ref{s:geo}. We assume here an intermediate value of the
covering factor, $f_{\rm c}=0.1$ (see Fig.\ 6), for which $kT=52^{+7}_{-5}$
keV, $\tau=0.82^{+0.10}_{-0.12}$ ($\chi^2=45/79$ d.o.f.), implying $y=0.33$.
Equating the fitted $\tau$ to $\tau_{\rm max}$, we obtain $\alpha\simeq 1$,
which is relatively large [as it appears to be generally the case in luminous
black hole systems (Narayan 1996)]. Fig.\ 8 shows the relation between $\tau$
and $\dot m$ ($\equiv \dot M c^2 /L_{\rm E}$) for those $y$ and $\alpha$; we
see that $\dot m_{\rm max}\simeq 4$.

\begin{figure} \begin{center} \leavevmode \epsfxsize=6.2cm
\epsfbox{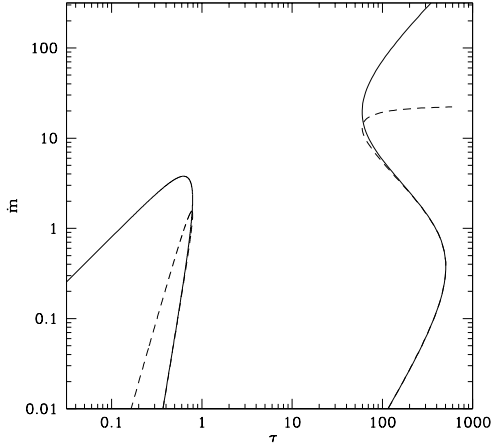}
\end{center} \caption{Accretion disc solutions at $r=13$. The optically-thin
solution (Z98) is given for $y=0.33$, and $\alpha=1$. We identify the observed
state of GX 339--4 with the maximum of this solution. For comparison, we also
show the optically-thick solution (Shakura \& Sunyaev 1973; Abramowicz et al.\
1988) computed for the same $\alpha$ and $M=3M_\odot$. Solid curves give the
accreted $\dot m$ whereas the dashed curves give the part of $\dot m$ which is
locally converted into escaping radiation. The remaining part of the
locally-dissipated energy is advected into the black hole. } \end{figure}

The maximum possible luminosity from the hot flow corresponds to local
accretion rate at the maximum set by advection (Fig.\ 8), for which radial
integration implies,
\begin{equation} \label{L_max}
L_{\rm hot}\la 0.15 y^{3/5} \alpha^{7/5} L_{\rm E} \end{equation}
(Z98). For the luminosity in Table 2, this relation yields $M_{\rm X}\ga
2.5M_\odot$, which is compatible with our estimates of the black hole mass in
Section 2. E.g.\ $L_{\rm hot}\simeq 0.06 L_{\rm E}$ at $M_{\rm X}=3M_\odot$.

Based on our data for GX 339--4 alone, we cannot distinguish which branch of
the optically thin solution (Fig.\ 8) the source follows in the hard state at
luminosities lower than the maximum one. In the $\alpha$ model, the advective
branch is stable (e.g.\ Wu 1997) whereas the cooling-dominated branch is
unstable (Pringle 1976), but the validity of those predictions has not been
tested observationally yet. We point out that if the flux variability in the
hard state within a factor of $\sim 50$ (see U94) corresponds to the advective
branch of the disc solution, the average accretion rate in the hot flow is
within a factor of $\sim 2$ of $\mmax$ (compare the left-hand dashed and solid
curves in Fig.\ 8). The corresponding large $\langle \dot M\rangle$ requires
then the presence of a radiation-driven outflow from the companion star for our
derived system parameters, see Section 2.3.

The formation of the hot disc might be related to the instability of the cold
disc in the inner part dominated by radiation pressure (Lightman \& Eardley
1974; Shakura \& Sunyaev 1976). If this is the case, the transition radius is
given by, \begin{equation} \label{eq:trans} {r_{\rm tr}\over J(r_{\rm
tr})^{16/21} } \approx 60 (\alpha M_{\rm X}/M_\odot)^{2/21} \dot m^{16/21}\,,
\end{equation} (e.g.\ Svensson \& Zdziarski 1994), where  $\alpha$ is the
viscosity parameter in the cold disc, not necessarily the same as that in the
hot one. For $\dot m\simeq 4$ and $\alpha\simeq 1$, $r_{\rm tr}\sim 200$.

Our best-fit  model has $L_{\rm bb}/L\simeq 0.3$, see Section \ref{s:soft}. If
the flow is on average $\sim 50$ per cent advective, $L_{\rm bb}/L \sim
50/r_{\rm tr}$ ($r_{\rm tr}\gg 50$), which then implies $r_{\rm tr}\sim 150$,
in good agreement with equation (\ref{eq:trans}) above. Then, $R_{\rm
tr}=6\times 10^7$ cm obtained in Section \ref{s:soft} corresponds to $M_{\rm
X}\sim 3$, in good agreement with other estimates of $M_{\rm X}$ here.

\subsection{The physical state of the reflecting medium}
\label{s:refl}

The region of formation of the Fe K$\alpha$ line is strongly ionized (Section
4, see also U94). At the best-fit ionization parameters obtained in Table 2, we
find that the dominant Fe ions are Fe\,{\sc xix} and Fe\,{\sc xx} for the first
and second observation, respectively. Those ions cannot, however, produce the
observed line because of the very strong resonant absorption (Matt, Fabian \&
Ross 1993; \.Zycki \& Czerny 1994). The observed line is produced by the
nonresonant Fe ions $\leq$ {\sc xvi}. We have computed that those ions
constitute about 20 per cent of all iron in the reflecting medium, which
explains the weakness of the line. The observed line equivalent width of $\sim
40$ eV is then fully consistent with theoretical calculations for $\af\sim 3$,
$\Omega/2\pi\simeq 0.4$ and the obtained ionization state of the medium, see
\.Zycki \& Czerny (1994), George \& Fabian (1991) and Matt et al.\ (1993). That
we did not require redshifts of the line center energy nor intrinsic line
broadening is consistent with the line origin at large distances from the black
hole, $r\gg 1$.

We can also obtain an estimate of the transition radius from the ionization
state of the reflector. The density, $n$, corresponding to unit Thomson optical
depth (where most of Compton reflection takes place) can be calculated from the
vertical structure of the standard cold disc (Milson, Chen \& Taam 1994). E.g.\
for $M_{\rm X}=3M_\odot$ and $\dot m=3$, $n=1.8\times 10^{20}$ cm$^{-3}$,
$6.0\times 10^{19}$ cm$^{-3}$, $1.8\times 10^{19}$ cm$^{- 3}$, and $1.9\times
10^{17}$ cm$^{-3}$ at $r=20$, 60, 200, and 2000, respectively. If there is a
central hot disc irradiating the outer cold disc, we can estimate the distance
between the illuminating source and the disc region where most of reflection
takes place to roughly equal the transition radius. The illuminating luminosity
on each side of the disc is $\sim (1/2) (\Omega/2\pi) L_{\rm hot}$. From that
we obtain $\xi\simeq 1200$, 500, 120, 120 erg cm s$^{-1}$ at the 4 values of
$r$ above, respectively. Thus, we see that $r_{\rm tr}\ga 10^2$ is compatible
with the fitted values of $\xi$ (see Table 2).

On the other hand, the illuminating luminosity is just $L_{\rm hot}$ in the
model of an active region above the disc (Fig.\ 5c). From energy balance, we
found such an active region is located at the height corresponding to its size
(Section \ref{s:geo}), and this height has to be $<R$. Then, using the values
of density from above, we find $\xi\ga 2000$ erg cm s$^{-1}$ at $r\la 100$.
Thus, the disc ionization expected in the active region model is much more than
that observed. If there are $k>1$ active regions with a typical luminosity of
$L_{\rm hot}/k$, its typical size (and thus, the height) must be at least
$<R/k^{1/2}$ to fit on the disc surface, and thus the above estimate of $\xi$
remains unchanged. This estimate provides one more argument against the active
region model. We caution, however, that the above results have been obtained
using our highly simplified model of photoionization (see Done et al.\ 1992).

Our conclusions here differ from those of U94, who estimated that reflection
originates in a region at $r\sim 10$. The difference arises because U94 used
the average disc density (which is $\gg$ the photosphere density) as well as
underestimated $\dot m$ by using the luminosity in the $\ginga$ energy range
only and neglecting advection.

\subsection{The role of magnetic fields}
\label{s:magn}

So far, we have neglected in our treatment any magnetic fields, which may be
present in GX 339--4 (Fabian et al.\ 1982; Di Matteo, Celotti \& Fabian 1997).
If present, thermal synchrotron radiation will provide additional seed photons
for thermal Comptonization (Zdziarski 1985, 1986), and this process has been
suggested to play a major role in luminous black-hole sources (e.g.\ Narayan
1996; Di Matteo et al.\ 1997). For $kT$ and $\tau$ typical of compact objects,
the synchrotron radiation is self-absorbed up to an energy, $E_s$,
corresponding to a high harmonic. The emitted spectrum is of the Rayleigh-Jeans
form below $E_s$, and exponentially cut off above it (Petrosian 1981; Takahara
\& Tsuruta 1982). Zdziarski (1986) has derived an expression for $E_s$ without
taking into account the angular dependence of the synchrotron radiation. Here,
we modify his expression to take into account the effect of angular averaging
as calculated by Mahadevan, Narayan \& Yi (1996) to obtain,
\begin{equation}\label{eq:es} \epsilon_{\rm s}={343\over 36}\Theta^2
\epsilon_{\rm c} \ln^3{C\over \ln{C\over \ln{C\over \dots}}}, \end{equation}
where $\epsilon_{\rm s}\equiv E_s/m_{\rm e} c^2$, $\epsilon_{\rm c}=B/B_{\rm
cr}$, the critical magnetic field strength is $B_{\rm cr}= m_{\rm e}^2 c^3/
e\hbar\simeq 4.4\times 10^{13}$ G,
\begin{equation}\label{eq:C} C={3\over 7\Theta} \left[ \pi \tau A_{\rm M}
\exp(1/\Theta) \over 3\alpha_{\rm f} \epsilon_{\rm c} \right]^{2/7},
\end{equation}
$\alpha_{\rm f}$ is the fine-structure constant and $A_{\rm M}(\Theta,
\epsilon_{\rm s}/\epsilon_{\rm c})$ is the low-energy and low-temperature
correction defined here as the ratio of equations (33) to (31) in Mahadevan et
al.\ (1996) with coefficients given in their Table 1. The factor $C$ above is
given in the limit of $\Theta\ll 1$, appropriate for GX 339--4.

In general, any magnetic field in the source will have the strength less than
that corresponding to pressure equipartion. In a hot accretion disc, the
largest contribution to pressure comes from hot ions (Section
\ref{s:accretion}), and thus $B^2/24\pi \la n kT_{\rm i}$. The ion temperature
in the disc is found to be always sub-virial, $T_{\rm i} \la 2\times 10^{11}$ K
(e.g.\ S76; Chen et al.\ 1997). The strongest magnetic field is achieved in an
inner region, where the plasma density is the largest. For estimates, we take a
region with the size of $R\sim 15GM_{\rm X}/c^2$, where dissipation per unit
logarithmic radius is maximized. In a disc-like geometry, hydrostatic
equilibrium implies $H/R\simeq 1$ around $\dot M_{\rm max}$ (e.g.\ Z98). For
$\tau\sim 1$ fitted in the slab geometry (Sections \ref{s:geo}--\ref{s:soft}),
equipartition corresponds then to $B\simeq 2\times 10^7$ G. Equations
(\ref{eq:es})-(\ref{eq:C}) then yield $E_s\simeq 10$ eV, or $\epsilon_{\rm s}
\simeq 2\times 10^{-5}$. ($A_{\rm M}\sim 0.1$ for the parameters used above.)

Photons in the resulting self-absorbed spectrum undergo Compton upscattering in
the thermal plasma. The spectral index of the Comptonized emission, $\Gamma
\simeq 1.75$ (see Table 2) implies certain ratio between the Comptonized
luminosity to that in the seed photons (see Zdziarski 1986). Assuming the
source area of $2\pi R^2$, the Comptonized-synchrotron luminosity is
approximately,
\begin{equation} \label{eq:lcs} L_{\rm CS}\simeq { 10 m_{\rm
e} c^3 \over \lambda^3 } {(2\Theta)^{3-\Gamma}\over 2-\Gamma} \epsilon_{\rm
s}^{\Gamma+1} R^2 \,, \end{equation}
where $\lambda \simeq 2.43\times 10^{-10}$ cm is the electron Compton
wavelength. For the source parameters obtained here, $L_{\rm CS} \simeq 3\times
10^{34}$ erg s$^{-1}$. This is 3 orders of magnitude below the
observed X\g\ luminosity of $\sim 3\times 10^{37}$ erg s$^{-1}$. Thus, our
conclusion is that the process of thermal synchrotron emission is negligible
for the formation of the X\g\ spectrum provided the emitting region is similar
in size to the region where the accretion energy is dissipated. Di Matteo et
al.\ (1997) have obtained a larger $L_{\rm CS}$ in GX 339--4 due to their
adoption of $kT \simeq 80$ keV, a much stronger $B$-field than that derived
above, and neglecting a correction to the rate of the synchrotron emission at
low $kT$ [as given in Table 1 of Mahadevan et al.\ (1996)]. Still, $L_{\rm
CS}\sim L$ for the observations considered here would require a source size
much larger than the region where gravitational energy dissipation is efficient
and/or magnetic field much above equipartition (equations
\ref{eq:es}--\ref{eq:lcs}).

We also note that equations above imply an approximate scaling of $L_{\rm
CS}/L_{\rm E} \propto M_{\rm X}^{(1-\Gamma)/2}$. Thus, Comptonized thermal
synchrotron emission from inner hot accretion discs around supermassive black
holes in Seyfert AGNs (with typical $\Gamma\sim 1.9$, $\Theta\sim 0.2$,
$\tau\sim 1$, e.g., Z97) will be even less important than in black-hole
binaries.

\subsection{$\bmath{e^\pm}$ pair production}
\label{s:pair}

If the hot plasma is purely thermal, its parameters of $kT\sim 50$ keV and
$\tau\sim 1$ imply that \ee\ pair production is due to photons in the extreme
Wien tail of the thermal spectrum, and thus it is very inefficient. Indeed,
performing standard calculations of the pair production rate using the model
spectrum as in Table 2 for the data set 1 and balancing that rate against the
pair annihilation rate (e.g.\ Svensson 1984) leads us to the conclusion that
there are almost no \ee\ pairs in the plasma. Specifically, the compactness,
$\ell\equiv L_{\rm hot}\sigma_{\rm T}/R m_{\rm e} c^3$, required for pair
production to be able to produce pairs with the fitted $\tau$ is $\sim 10^4$
for the hot source modelled as a sphere with radius $R$. This corresponds to
$R\sim 10^5$ cm, i.e., much less than even the Schwarzschild radius of the
black hole.

We have also performed pair-balance calculations for a slab, which better
approximates the disc geometry. Here, we used the model of a slab with cold
clouds in the midplane, see Fig.\ 5d and Section \ref{s:geo}. The relevant
compactness is then the local one corresponding to the power dissipated within
a cube with size equal to the half-thickness, $H$, of the slab. The results are
shown in Fig.\ 6, where we see that $\ell\sim 10^3$--$3\times 10^3$ (depending
on the covering factor of cold clouds, $f_{\rm c}$). Approximating the source
as a uniformly radiating disc with the outer radius $R$, we obtain
$R^2/H=L_{\rm hot}\sigma_{\rm T}/2\pi\ell m_{\rm e} c^3$, which for $H/R\sim 1$
characteristic for hot discs yields $R\sim 10^5$ cm, which is the same value as
obtained above for the spherical geometry, and much less than our estimates of
the size of the hot plasma.

Thus, there are no \ee\ in the hot plasma in GX 339--4 provided the plasma is
fully thermal. On the other hand, it is also possible that nonthermal
acceleration of selected electrons to relativistic energies operates in the
source in addition to heating of the plasma. This hybrid model has been applied
to NGC 4151 by Zdziarski et al.\ (1996) and Johnson et al.\ (1997) and to Cyg
X-1 by Poutanen \& Coppi (1998) and Poutanen (1998). We have applied that model
to the data set 1. The pair-dominated hybrid model yields the same $\chi^2$ as
the thermal model (fit 1c in Table 2) for any $\ell \ga 150$, and the 90 per
cent confidence lower limit is $\ell_{\rm min}\simeq 70$ (for spherical
geometry). The fraction of the power that is supplied to the accelerated
electrons and pairs is $\simeq 0.09$ at $\ell_{\rm min}$, and it decreases with
increasing $\ell$. This $\ell_{\rm min}$ implies $R< 10^7$ cm, which
corresponds to $r\la 20$ for $M_{\rm X}=3M_\odot$. This is in principle
compatible with the expected size of a hot accretion disc ($r\gg 6$).

We have also applied the hybrid model to the data obtained from the entire OSSE
observation of the source, 1991 September 5--12 (G95), which have much better
statistical quality than either of our OSSE data sets. We find those data are
fitted somewhat better, $\Delta\chi^2=-4$, by the hybrid model than by the pure
thermal one (which gives $kT=62^{+13}_{-9}$ keV at fixed $\Omega/2\pi=0.44$).
The residuals in the thermal model show a weak high-energy tail on top of the
thermal spectrum above $\sim 300$ keV. This may hint for the presence of
nonthermal pairs, accounting for the tail. In the hybrid model, $\ell=
150^{+\infty}_{-30}$, and the nonthermal fraction at the best fit is $\simeq
0.05$. On the other hand, G95 show softening of the spectrum of GX 339--4 with
time during that observation. Therefore, the tail could also be an artefact of
fitting the spectrum averaged over a range of plasma parameter by a
single-component thermal model.

\section{COMPARISON WITH BLACK-HOLE AND NEUTRON-STAR SOURCES}

In this section, we compare our X\g\ spectrum of GX 339--4 with those of
established black-hole and neutron-star sources. Our objective is to determine
whether the X\g\ spectrum of GX 339--4 is indeed similar to those seen in
black-hole sources but not to those of accreting neutron stars. We stress that
both black-hole binaries and neutron-star binaries exhibit two main spectral
states, X-ray low (hard) and high (soft), and only the corresponding states
should be directly compared.

\subsection{X$\bmath{\gamma}$ emission of black-hole sources}

\subsubsection{Black-hole binaries in the hard state}

Hard-state X-ray spectra of the archetypical black-hole binary Cyg X-1 have
been fitted by G97. Their fit to the average \ginga\/ data of 1991 June 6 with
a power-law and reflection model (as in Section 4) yields $\Gamma=
1.59^{+0.03}_{-0.03}$ and $\Omega/2\pi= 0.34^{+0.05}_{-0.05}$, which is rather
similar to our spectra of GX 339--4. X-ray spectra with similar power-law
indices and moderately weak Compton-reflection components are seen from other
black-hole binaries in the hard (low) state, e.g.\ Nova Muscae (Ebisawa et al.\
1994; \.Zycki et al.\ 1998) and GS 2023+338 (\.Zycki et al.\ 1997).

The broad-band X\g\ spectra of Cyg X-1 in G97 are also similar to that of GX
339--4, and well modelled by a primary continuum due to thermal Comptonization
in a plasma with similar $\tau$, but higher electron temperature, $kT\sim 100$
keV (and with an additional small spectral component from emission of a plasma
with $\tau\gg 1$). Soft \g-ray spectra similar in shape to that of Cyg X-1 have
been observed from other black-hole binaries in the hard state (e.g.\ Grove et
al.\ 1998; Grebenev, Sunyaev \& Pavlinsky 1997).

Motivated by the similarity, we test here the hot accretion disc model of Z98
(see Section \ref{s:accretion}) against the Cyg X-1 data. We have refitted the
spectrum of G97 corresponding to the peak flux using the hot slab model used in
Section \ref{s:geo} for the primary continuum, and obtained $y\simeq 0.4$ and
$\tau\simeq 0.5$ (assuming $f_{\rm c}=0.1$). Those parameters imply
$\alpha\simeq 0.4$ using $\tau(\mmax)$ of equation (6). The peak luminosity in
the hard state is about $4\times 10^{37}$ erg s$^{-1}$, for which equation (7)
implies $M_{\rm X}\ga 10M_\odot$. This agrees well with best estimates of the
mass of Cyg X-1 (e.g.\ van Paradijs \& McClintock 1995). Thus, it is possible
that a hot accretion disc is present in the hard state of black-hole binaries,
and that its structure determines both the observed optical depth of the hot
plasma and the maximum luminosity in the hard state.

\subsubsection{Seyfert 1s}

\begin{figure}
\begin{center}
\leavevmode
\epsfxsize=7cm \epsfbox{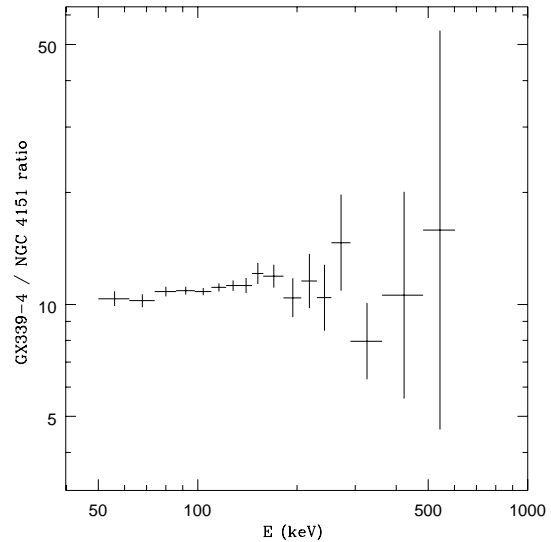}
\end{center}
\caption{The ratio of the OSSE spectrum of GX 339--4 from 1991 September 5--12
to the average OSSE spectrum of the Seyfert NGC 4151.
 }
\end{figure}

It is of great interest that the X\g\ spectra of black-hole binaries in the
hard state are similar to the spectra of Seyfert-1 AGNs, which similarity in
the X-ray regime was pointed out by Tanaka (1989). Specifically, both classes
of objects show power-law X-ray spectra with $\Gamma \sim 1.7$--2, Compton
reflection components with Fe K$\alpha$ lines, and high-energy cutoffs above
$\sim 100$ keV (e.g.\  Z97). The cutoffs can be modelled in both cases by
Comptonization with $\tau\sim 1$ and $kT\sim 100$ keV (Z97).

Here, we point out that the intrinsic spectrum above 4 keV of GX 339--4 appears
virtually identical to that of NGC 4151, which, in soft \g-rays,  is the
brightest and best-studied Seyfert AGN. We note here that apart from strong
absorption, the X\g\ spectrum of NGC 4151 appears rather typical for Seyfert
1s, and the OSSE spectra of NGC 4151 and of all other Seyfert 1s observed by
OSSE are very similar (Zdziarski et al.\ 1996; Z97). Also, the intrinsic X\g\
spectrum of NGC 4151 is very similar to that of a `typical' Seyfert 1, NGC 5548
(Magdziarz et al.\ 1998). This makes models specially designed for NGC 4151
(e.g.\ Poutanen et al.\ 1996) probably unnecessary. The thermal Comptonization
fit (after applying the same correction to solutions of the Kompaneets equation
as in Section 4 above) to a joint {\it ROSAT}/\ginga/OSSE spectrum of NGC 4151
(Zdziarski et al.\ 1996) yields $\Gamma= 1.81^{+0.04}_{-0.04}$, $kT=
63^{+19}_{-11}$ keV , $\Omega/2\pi =0.49^{+0.26}_{-0.26}$, which are consistent
within errors with those given in Table 2 for GX 339--4. We find that the form
of the high-energy cutoff is almost the same in both sources, as shown in Fig.\
9, which shows the ratio of the OSSE spectrum of GX 339--4 (from the entire
observation of 1991 September 5--12) to the spectrum of NGC 4151 averaged over
all OSSE observations (from Johnson et al.\ 1997).

\subsection{X$\bmath{\gamma}$ emission of weakly-magnetized
accreting neutron-stars}

Two types of objects whose X\g\ emission (if present) is unambiguously
connected with neutron stars are X-ray pulsars and isolated neutron stars.
Those X\g\ sources possess strong magnetic fields, $B\ga 10^{11}$ G, and their
emission can be easily distinguished from that of black-hole systems, see
Finger \& Prince (1997) and Thompson et al.\ (1997) for recent reviews.

On the other hand, weakly-magnetized accreting neutron stars often show X-ray
emission very similar to that of black-hole binaries. Some LMXBs classified as
black-hole candidates have been subsequently found to emit type-1 X-ray bursts,
leading to their identifications as neutron-star binaries. A recent example is
discovery of X-ray bursts from GS 1826--238 (Bazzano et al.\ 1997), which
source appeared as a black-hole candidate in Tanaka (1989).

Indeed, X-ray emission of X-ray bursters in their hard (low) state can be very
similar to that of GX 339--4. For example, two \ginga\/ observations of 4U
1608--522 show a power law component with $\Gamma\simeq 1.9$ accompanied by
Compton reflection from a strongly ionized medium with $\Omega/2\pi\sim 0.5$
in the $\sim 2$--60 keV range (Yoshida et al.\ 1993). Similarly, GS 1826--238
observed by \ginga\/ has $\Gamma\sim 1.8$ and Compton reflection (Strickman et
al.\ 1996). Thus, X-ray spectra alone appear not sufficient to distinguish a
weakly-magnetized neutron star from a black hole.

On the other hand, typical spectra of X-ray bursters observed in the $\sim
30$--200 keV range by SIGMA telescope on board {\it GRANAT\/} and the BATSE
detector on board {\it CGRO} are rather soft, see reviews by Barret \& Vedrenne
(1994), van der Klis (1994), Vargas et al.\ (1997) and Tavani \& Barret (1997).
In most cases, the spectra can be fitted equally well by a power law with
$\langle \Gamma\rangle \sim 3$ or thermal Comptonization with $\langle
kT\rangle \sim 15$--20 keV. There is only one reported OSSE detection of an
X-ray burster, GS 1826--238, which shows a similar $\Gamma=3.1\pm 0.5$ above 50
keV (Strickman et al.\ 1996). These spectra are all much softer than the
corresponding spectra of black hole binaries in the hard state [but note their
similarity to black-hole binary spectra in the soft state (Grove et al.\
1998)].

Unfortunately, there have been almost no observations below 30 keV simultaneous
with those at $\ga 30$ keV discussed above. Therefore, there is no information
on the X-ray spectral state (high or low) of the bursters during most of those
observations. However, X-ray bursters are rather often found in the low state
when observed in X-rays alone, and thus we can assume that some of the spectra
observed at $\ga 30$ keV correspond to the X-ray low, hard, state. Then, this
implies that the X\g\ spectra of bursters in the low state have spectral breaks
at energies significantly below 100 keV, or have thermal-Comptonization
temperatures of $kT\la 30$ keV. This conclusion is indeed confirmed by a $\sim
2$--250 keV observation of 4U 1608--522 in the low state simultaneously by
\ginga\/ and BATSE, which yields a break at $\sim 60$ keV in the broken
power-law model or $kT\sim 25$ keV, $\tau\gg 1$, in the thermal Comptonization
model (Zhang et al.\ 1996). A similar spectral break at $\sim 60$ keV was also
seen in a 30--200 keV observation of 4U 1728--34 (Claret et al.\ 1994).

\subsection{Spectral black-hole signatures}

As discussed above, X-ray bursters in the low state have X\g\ emission with
breaks or cutoffs at significantly {\it lower\/} energies than both that seen
in GX 339--4 and those found in black-hole binaries in the hard (low) state in
general (Section 6.1.1). When the spectra are fitted by the
thermal-Comptonization model (with $\tau\ga 1$), $kT\la 30$ keV and $kT\ga 50$
keV for neutron stars and black holes, respectively. We propose this
quantitative criterion to be a black-hole signature.

We see that these two ranges of temperature are not far apart, and it is
important to test this criterion against available data. Those tests should
take into account Compton reflection and relativistic effects in thermal
Comptonization, and use broad-band X\g\ data. These criteria are not satisfied
by thermal-Comptonization fits in Mandrou et al.\ (1994) and Churazov et al.\
(1994), who find $kT= 33$ keV and 38 keV, respectively, in GRS 1758--258, a
black-hole candidate in the hard state (Sunyaev et al.\ 1991). Those fits use
the nonrelativistic model of Sunyaev \& Titarchuk (1980), neglect reflection,
and are in a 35--250 keV range only. In fact, very similar assumptions lead to
a gross underestimate of the temperature in Cyg X-1 in the hard state, $kT=27$
keV (Sunyaev \& Tr\"umper 1979), whereas $kT\sim 100$ keV is obtained from
broad-band spectra fitted by a relativistic Comptonization model with
reflection (e.g.\ G97).

As a caveat, we mention that a 40--200 keV SIGMA spectrum of a source in Terzan
2 globular cluster appears to have no cutoff up to 200 keV and $\Gamma\simeq
1.7\pm 0.5$ (Barret \& Vedrenne 1994; Barret et al.\ 1991). Such a spectrum may
imply a Comptonization temperature $\ga 100$ keV, which is characteristic of
black-hole sources but not of neutron-star ones, whereas the same cluster
contains an X-ray burster, X1724--308.

Finally, we point out that the presence of power-law emission with $\Gamma\sim
2.5$--3 in the $\sim 30$--500 keV range, which is seen from black hole binaries
in the soft state (Grove et al.\ 1998), may not constitute a black-hole
signature, in spite of a recent claim (Titarchuk 1997). Very similar soft
single power-law spectra without detectable high-energy breaks are seen from
many X-ray bursters (e.g.\ Barret et al.\ 1992; Goldwurm et al.\ 1996; Harmon
et al.\ 1996).

\section{CONCLUSIONS}

The results of this work can be divided into two main parts as far as
dependence on model assumptions is concerned. The results of Sections 2, 3, 4
and 6 are relatively model-independent.

In Section 2, we determine the distance to the object to be $>3$ kpc, with 4
kpc being the most likely value. We show the most recent distance determination
of $\sim 1.3$ kpc is in error. We also find $\ebv\simeq 1.2$ and $\nh\simeq
6\times 10^{21}$ cm$^{-2}$ as mutually-consistent most likely values. The mass
of the compact object appears relatively low, $M_{\rm X}\la 5 M_\odot$.

In Sections 3 and 4, we present our data and show that the spectra can be very
well fitted by Comptonization in a thermal plasma with $kT\simeq 50$ keV and
$\tau\sim 1$. In addition, we find, at very high significance, the presence
of Compton reflection from an ionized medium and a soft X-ray excess.

In Section 6, we show that this spectrum is similar to that of other black-hole
binaries, as well as to those of Seyfert AGNs. After comparison with spectra of
neutron-star sources, we propose that a thermal-Comptonization temperature of
$kT\ga 50$ keV represents a black-hole signature.

Physical interpretation of our results is given in Section 5. Here, we study
constraints following from the X\g\ spectra and concentrate on accretion
disc and corona models. We do not consider, e.g.\ models with large scale
outflows or jets. Since the X\g\ observations were not accompanied by
observations in any other wavelength, we do not discuss such data (e.g.\
optical, Motch et al.\ 1983) obtained at other epochs. We also do not discuss
time variability of the source.

With these caveats, our best physical model is that of a hot accretion disc
within $\sim 100$ gravitational radii surrounded by a cold outer disc, see
Figs.\ 5d, e. The seed photons for thermal Comptonization in the hot disc are
supplied by cold clouds within the hot disc. The emission of the hot disc is
Compton-reflected by the outer cold disc. The outer disc also emits most of the
observed soft X-ray excess. This model is in agreement with the spectra, energy
balance, and ionization balance at the surface of the reflecting outer disc.
The observed amount of reflection requires that the outer disc is flared.

The hot-disc accretion rate is near the maximum set by advection. Based on the
spectral fit of the hot slab model, we find the viscosity parameter of
$\alpha\sim 1$ and $M_{\rm X}\ga 3M_\odot$, which mass is in agreement with the
dynamical mass determination. The hot disc model, which parameters are
independent of $M_{\rm X}$,  is also supported by the observed similarity of
the spectrum of GX 339--4 to those of Seyfert 1s.

We find that \ee\ pair production photons in the thermal-Comptonization
spectrum is negligible and thus the disc is most likely made of electrons and
ions (although more complex models with \ee\ pairs are possible). Also,
synchrotron emission in the hot disc with equipartition magnetic field is
negligible as a source of seed photons for Comptonization.

We can rule out models with a cold disc covered by a homogeneous corona or by
active regions located on the surface of the disc as violating the energy
balance. On the other hand, the energy balance is satisfied if there are active
regions at some height above the disc surface. However, this model provides
poor fits to the spectral data, the predicted ionization state of the
postulated inner cold disc is much higher than that found from the
Compton-reflection spectral component, as well as it predicts an
$\Omega$-$\Gamma$ correlation opposite to that observed.

\section*{ACKNOWLEDGMENTS}

This research has been supported in part by the KBN grants 2P03C00511p0(1,4),
2P03D01008 and 2P03D00624, NASA grants and contracts, the Swedish Natural
Science Research Council, Stockholm University, and the Anna-Greta and Holger
Crafoord's Fund.  We thank Pawe\l\ Magdziarz for his assistance with
implementing models into the {\sc xspec} software package, and Bo\.zena Czerny,
Agata R\'o\.za\'nska and Grzegorz Wardzi\'nski for valuable discussions.

\bsp

\label{lastpage}


\begin{thebibliography}{}

\bibitem[]{}
Abramowicz M. A., Czerny B., Lasota J.-P., Szuszkiewicz E., 1988, ApJ, 332,
646

\bibitem[]{}
Abramowicz M. A., Chen X., Kato S., Lasota J.-P., Regev O., 1995, ApJ, 438,
L37

\bibitem[]{}
Alcaino G., 1971, A\&A, 11, 7

\bibitem[]{}
Anders E., Ebihara M., 1982, Geochim.\ Cosmochim.\ Acta, 46, 2363

\bibitem[]{}
Arnaud K. A., 1996, in Jacoby G. H., Barnes J., eds., Astronomical Data
Analysis Software and Systems V, ASP Conf. Series Vol.\ 101, San Francisco,
p.\ 17

\bibitem[]{}
Barret D., Vedrenne G., 1994, ApJS, 92, 505

\bibitem[]{}
Barret D., et al., 1991, ApJ, 379, L21

\bibitem[]{}
Barret D., et al., 1992, ApJ, 394, 615

\bibitem[]{}
Bazzano A., Cocchi M., Natalucci L., Ubertini P., Heise J., in't
Zand J., Muller J. M., Smith M. J. S., 1997, in Dermer C. D., Strickman M. S.,
Kurfess J. D., eds, The 4th Compton Symposium, AIP, New York, p.\ 729

%\bibitem[]{}
%Ba\l uci\'nska-Church M., Belloni T., Church M. J., Hasinger G., 1995,
%A\&A, 302, L5

%\bibitem[]{}
%Bj\"ornsson G., Abramowicz M. A., Chen X., Lasota J.-P., 1996, ApJ, 467, 99

\bibitem[]{}
Burton W. B., 1992, in Burton W. B., Elmgreen B. G, Genzel R., eds., The
Galactic Interstellar Medium, Springer, Berlin, p.\ 53

\bibitem[]{}
Callanan P. J., Charles P. A., Honey W. B., Thorstensen J. R., 1992, MNRAS,
259, 395 (C92)

%\bibitem[]{}
%Chen K., Halpern J. P., 1989, ApJ, 344, 115

\bibitem[]{}
Celotti A., Fabian A. C., Rees M. J., 1992, MNRAS, 255, 419

\bibitem[]{}
Chen X., Abramowicz M. A., Lasota J. P., 1997, ApJ, 476, 61

\bibitem[]{}
Churazov, E., et al., 1994, ApJS, 92, 381

\bibitem[]{}
Claret A., et al.\ 1994, ApJ, 423, 436

\bibitem[]{}
Collin-Souffrin S., Czerny B., Dumont A.-M., \.Zycki, P.~T., 1996,
A\&A, 314, 393

\bibitem[]{}
Coppi P. S., 1992, MNRAS, 258, 657

\bibitem[]{}
Corbet R. H. D., Thorstensen J. R., Charles P. A., Honey W. B.,
Smale A. P., Menzies J. W., 1987, MNRAS, 227, 1055

\bibitem[]{}
Cowley A. P., Crampton D., Hutchings J. B., 1987, AJ, 92, 195 (C87)

\bibitem[]{}
Di Matteo T., Celotti A., Fabian A. C., 1997, MNRAS, 291, 705

\bibitem[]{}
Diplas A., Savage B. D., 1994, ApJ, 427, 274

\bibitem[]{}
Done C., Mulchaey J. S., Mushotzky R. F., Arnaud K. A., 1992, ApJ, 395, 275

\bibitem[]{}
Doxsey R., et al., 1979, ApJ, 228, L67

\bibitem[]{}
Ebisawa K., et al., 1994, PASJ, 46, 375

\bibitem[]{}
Ebisawa K., Ueda Y., Inoue H., Tanaka Y., White N. E., 1996, ApJ, 467, 419

\bibitem[]{}
Fabian A. C., Guilbert P. W., Motch C., Ricketts M., Ilovaisky S. A.,
Chevalier C., 1982, A\&A, 111, L9

\bibitem[]{}
Fabian A. C., Rees M. J., Stella L., White N. E., 1989, MNRAS, 238, 729

%\bibitem[]{}
%Fabian A. C., Nandra K., Reynolds C. S., Brandt W. N., Otani C., Tanaka Y.,
%1995, MNRAS, 277, L11

\bibitem[]{}
Finger M. H., Prince T. A., 1997, in Dermer C. D., Strickman M. S.,
Kurfess J. D., eds, The 4th Compton Symposium, AIP, New York, p.\ 57

\bibitem[]{}
FitzGerald M. P., 1987, MNRAS, 229, 227

\bibitem[]{}
FitzGerald M. P., Jackson P. D., Moffat A. F. J., 1977, Observatory, 97, 129

\bibitem[]{}
Galeev A. A., Rosner R., Vaiana G. S., 1979,  ApJ, 229, 318

\bibitem[]{}
George I. M., Fabian A. C., 1991, MNRAS, 249, 352

%\bibitem[]{}
%Georgelin Y. M., Georgelin Y. P., 1976, A\&A, 49, 57

\bibitem[]{}
Gierli\'nski M., Zdziarski A. A., Done C., Johnson W. N., Ebisawa K., Ueda Y.,
Haardt F., Phlips B. F., 1997, MNRAS, 288, 958 (G97)

\bibitem[]{}
Goldwurm A., et al., 1996, A\&A, 310, 857

\bibitem[]{}
Grabelsky D. A., et al., 1995, ApJ, 441, 800 (G95)

\bibitem[]{}
Grebenev S. A., Sunyaev R. A., Pavlinsky M. N., 1997, Adv.\ Sp.\ Res., 19,
(1)15

\bibitem[]{}
Grove J. E., Johnson W. N., Kroeger R. A., McNaron-Brown K., Skibo J. G.,
1998, ApJ, 500, 899

\bibitem[]{}
Grindlay J. E., 1979, ApJ, 232, L33

%\bibitem[]{}
%Haardt F., 1993, ApJ, 413, 680

\bibitem[]{}
Haardt F., Maraschi L., 1993, ApJ, 413, 507

\bibitem[]{}
Haardt F., Maraschi L., Ghisellini G., 1994, ApJ, 432, L95

\bibitem[]{}
Haensel P., 1995, in Roxburgh I. A., Masnou J.-L., eds., Physical Processes in
Astrophysics, Lecture Notes in Physics 458, Springer, Berlin, p.\ 49

\bibitem[]{}
Harmon B. A., et al., 1994, ApJ, 425, L17

\bibitem[]{}
Harmon B. A., Wilson C. A., Tavani M., Zhang S. N., Rubin B. C., Paciesas W.
S., Ford E. C., Kaaret P., 1996, A\&AS, 120, (III)197

\bibitem[]{}
Ichimaru S.,  1977, ApJ, 214, 840

\bibitem[]{}
Ilovaisky S. A., Chevalier C., 1981, IAU Circ.\ 3856

\bibitem[]{}
Ilovaisky S. A., Chevalier C., Motch C., Chiapetti L., 1986, A\&A, 164, 67

\bibitem[]{}
Johnson W. N., et al., 1993, ApJS, 86, 693

\bibitem[]{}
Johnson W. N., McNaron-Brown K., Kurfess J. D., Zdziarski A. A., Magdziarz
P., Gehrels N., 1997, ApJ, 482, 173

\bibitem[]{}
Kaastra J. S., Mewe R., 1993, A\&AS, 97, 443

\bibitem[]{}
Krolik J. H., 1998, ApJ, 498, L13

\bibitem[]{}
Kuncic Z., Celotti A., Rees, M. J., 1997, MNRAS, 284, 717

%\bibitem[]{} Lampton M., Margon B., Bowyer S., 1976, ApJ, 208, 177

\bibitem[]{} Lightman A. P., Eardley D. M., 1974, ApJ, 187, L1

\bibitem[]{} Lightman A. P., White T. R., 1988, ApJ, 335, 57

\bibitem[]{}
Lindoff U., 1972, A\&AS, 7, 231

%\bibitem[]{}
%Macio{\l}ek-Nied\'zwiecki A., Zdziarski A. A., Coppi P. S., 1995, MNRAS, 276,
%273

%\bibitem[]{}
%Maejima Y., Makishima K., Matsuoka M., Ogawara Y., Oda M., Tawara Y., Doi K.,
%1984, ApJ, 285, 712

\bibitem[]{}
Magdziarz P., Zdziarski A. A., 1995, MNRAS, 273, 837

\bibitem[]{}
Magdziarz P., Blaes O. M., Zdziarski A. A., Johnson W. N., Smith D. A., 1998,
MNRAS, in press

\bibitem[]{}
Mahadevan R., Narayan R., Yi I., 1996, ApJ, 465, 327

\bibitem[]{}
Makino F., the ASTRO-C team, 1987, Ap Let Com, 25, 223

\bibitem[]{}
Makishima K., et al., 1986, ApJ, 308, 635

\bibitem[]{}
Mandrou P., et al.\ 1994, ApJS, 92, 343

\bibitem[]{}
Matt G., Fabian A. C., Ross R. R., 1993, MNRAS, 262, 179

%\bibitem[]{}
%Matt G., Fabian A. C., Ross R. R., 1996, MNRAS, 278, 1111

\bibitem[]{}
Mauche C. W., Gorenstein P., 1986, ApJ, 302, 371

\bibitem[]{}
Mermilliod J.-C., Mermilliod M., 1994, Catalogue of Mean UBV Data
on Stars. Springer, Berlin

\bibitem[]{}
Milson J. A., Chen X., Taam R. E., 1994, ApJ, 421, 668

\bibitem[]{}
Miyamoto S., Kimura K., Kitamoto S., Dotani T., Ebisawa K., 1991, ApJ, 383,
784

\bibitem[]{}
Moffat A. F. J., Vogt N., 1973, A\&AS, 10, 135

\bibitem[]{}
Moffat A. F. J., Vogt N., 1975, A\&AS, 20, 155

\bibitem[]{}
Motch C., Ricketts M. J., Page C. G., Ilovaisky S. A., Chevalier C., 1983,
A\&A, 119, 171

\bibitem[]{}
Nandra K., Pounds K. A., 1994, MNRAS, 268, 405

\bibitem[]{}
Narayan R., 1996, ApJ, 462, 136

\bibitem[]{}
Narayan R., Yi I., 1995, ApJ, 452, 710

\bibitem[]{}
Neckel T., Klare G., 1980, A\&AS, 42, 251

%\bibitem[]{}
%Nowak M. A., 1995, PASP, 718, 1207

\bibitem[]{}
Petrosian V., 1981, ApJ, 251, 727

%\bibitem[]{}
%Piran T., 1978, ApJ, 221, 652

\bibitem[]{}
Podsiadlowski P., 1991, Nat, 350, 136

\bibitem[]{}
Poutanen J., 1998, in Abramowicz M. A., Bj\"ornsson G., Pringle J.
E., eds., Theory of Black Hole Accretion Discs, Cambridge Univ. Press,
Cambridge, in press (astro-ph/9805025)

\bibitem[]{}
Poutanen J., Coppi P. S., 1998, Physica Scripta, in press (astro-ph/9711316)

\bibitem[]{}
Poutanen J., Svensson R., 1996, ApJ, 470, 249

\bibitem[]{}
Poutanen J., Sikora M., Begelman M. C., Magdziarz P., 1996, ApJ, 465, L107

\bibitem[]{}
Poutanen J., Krolik J. H., Ryde F., 1997, MNRAS, 292, L21

\bibitem[]{}
Predehl P., Br\"auninger H., Burkert W., Schmitt J. H. M. M., 1991, A\&A, 246,
L40

\bibitem[]{}
Press W. H., Teukolsky S. A., Vetterling W. T., Flannery B. P., 1992, Numerical
Recipes. Cambridge Univ.\ Press, Cambridge

\bibitem[]{}
Pringle J. E., 1976, MNRAS, 177, 65

\bibitem[]{}
Reilman R. F., Manson S. T., 1979, ApJS, 40, 815

%\bibitem[]{}
%R\'o\.za\'nska A., Czerny B., \.Zycki P. T., 1998, in preparation

\bibitem[]{}
Rubin B. C., Harmon B. A., Paciesas W. S., Robinson C. R., Zhang S. N.,
Fishman G. J., 1998, ApJ, 492, L67

\bibitem[]{}
Rybicki G. R., Lightman A. P., 1979, Radiative Processes in Astrophysics.
Wiley-Interscience, New York

\bibitem[]{}
Shakura N. I., Sunyaev R. A., 1973, A\&A, 24, 337

\bibitem[]{}
Shakura N. I., Sunyaev R. A., 1976, MNRAS, 175, 613

\bibitem[]{}
Shapiro S. L., Lightman A. P., Eardley D. M., 1976, ApJ, 204, 187 (S76)

\bibitem[]{}
Scheffler H., Elsasser H., 1987, Physics of the Galaxy and Interstellar
Matter. Springer, Berlin

\bibitem[]{}
Shimura T., Takahara F., 1995, ApJ, 445, 780

\bibitem[]{} Smale A. P., 1996, in Evans A., Wood J. H., eds., Cataclysmic
Variables and Related Objects, Kluwer, Amsterdam, p.\ 347

\bibitem[]{}
Stern B. E., Poutanen J., Svensson R., Sikora M., Begelman M. C., 1995, ApJ,
449, L13

\bibitem[]{}
Strickman M., Skibo J., Purcell W., Barret D., Motch C., 1996, A\&AS, 120,
(III)217

\bibitem[]{}
Sunyaev R. A., Titarchuk L. G., 1980, A\&A, 86, 121

\bibitem[]{}
Sunyaev R. A., Tr\"umper J., 1979, Nat, 279, 506

\bibitem[]{}
Sunyaev R. A., et al., 1991, A\&A, 247, L29

\bibitem[]{}
Svensson R., 1984, MNRAS, 209, 175

\bibitem[]{}
Svensson R., Zdziarski A. A., 1994, ApJ, 436, 599

\bibitem[]{}
Takahara F., Tsuruta S., 1982, Prog.\ Theor.\ Phys., 67, 485

\bibitem[]{} Tanaka Y., 1989, in Hunt J., Battrick B., eds., 23rd ESLAB
Symposium, ESA SP-296, p.\ 3

\bibitem[]{}
Tanaka Y., Lewin W. H. G., 1995, in Lewin W. H. G., van Paradijs J., van den
Heuvel E. P. J., eds., X-Ray Binaries, Cambridge Univ.\ Press, Cambridge,
p.\ 126

\bibitem[]{}
Tavani M., Barret D., 1997, in Dermer C. D., Strickman M. S.,
Kurfess J. D., eds, The 4th Compton Symposium, AIP, New York, p.\ 75

\bibitem[]{}
Tavani M., London R., 1993, ApJ, 410, 281

\bibitem[]{} Titarchuk L., 1997, in Winkler C., Courvoisier T., Durouchoux P.,
eds., The Transparent Universe, ESA SP-382, p.\ 163

\bibitem[]{}
Thompson D. J, Harding A. K., Hermsen W., Ulmer M. P., 1997, in
Dermer C. D., Strickman M. S., Kurfess J. D., eds, The 4th Compton Symposium,
AIP, New York, p.\ 39

\bibitem[]{}
Turner M. J. L., et al., 1989, PASJ, 41, 345

%\bibitem[]{} Turner T. J., Done C., Mushotzky R. F., Madejski M., Kunieda H.,
%1992, ApJ, 391, 102

\bibitem[]{}
Ueda Y., Ebisawa K., Done C., 1994, PASJ, 46, 107 (U94)

\bibitem[]{}
van der Klis M., 1994, ApJS, 92, 511

\bibitem[]{}
van Paradijs J., McClintock J. E., 1995, in: Lewin W. H. G., van Paradijs J.,
van den Heuvel E. P. J., eds., X-Ray Binaries, Cambridge Univ.\ Press,
Cambridge, p.\ 58

\bibitem[]{}
Vargas M., et al., 1997, in Winkler C., Courvoisier T., Durouchoux
P., eds., The Transparent Universe, ESA SP-382, p. \ 129

\bibitem[]{}
V{\'a}zquez R. A., Feinstein A., 1992, A\&AS, 92, 863

\bibitem[]{}
Vrtilek S. D., Raymond J. C., Garcia M. R., Verbunt F., Hasinger G., K\"urster
M., 1990, A\&A, 235, 162

\bibitem[]{}
Vrtilek S. D., McClintock J. E., Seward F. D., Kahn S. M., Wargelin B. J.,
1991, ApJS, 76, 1127

\bibitem[]{}
Webbink R. F., Rappaport S., Savonije G. J., 1983, ApJ, 270, 678

\bibitem[]{}
Wu X.-B., 1997, MNRAS, 292, 113

\bibitem[]{}
Yoshida K, Mitsuda K, Ebisawa K., Ueda Y., Fujimoto R., Yaqoob T., Done C.,
1993, PASJ, 45, 605

\bibitem[]{}
Zdziarski A. A., 1985, ApJ, 289, 514

\bibitem[]{}
Zdziarski A. A., 1986, ApJ, 303, 94

\bibitem[]{}
Zdziarski A. A., 1998, MNRAS, 296, L51 (Z98)

%\bibitem[]{}
%Zdziarski A. A., Fabian A. C., Nandra K., Celotti A., Rees M. J., Done C.,
%Coppi P. S., Madejski G. M., 1994, MNRAS, 269, L55

\bibitem[]{}
Zdziarski A. A., Johnson W. N., Magdziarz P., 1996, MNRAS, 283, 193

\bibitem[]{}
Zdziarski A. A., Johnson W. N., Poutanen J., Magdziarz P., Gierli\'nski M.,
1997, in Winkler C., Courvoisier T., Durouchoux P., eds., The Transparent
Universe, ESA SP-382, p. \ 373 (Z97)

\bibitem[]{}
Zhang S. N., et al., 1996, A\&AS, 120, (III)279

\bibitem[]{}
\.Zycki P. T., Czerny B., 1994, MNRAS, 266, 653

\bibitem[]{}
\.Zycki P. T., Done C., Smith D. A., 1997, ApJ, 488, L113

\bibitem[]{}
\.Zycki P. T., Done C., Smith D. A., 1998, ApJ, 496, L25

\end{thebibliography}
\end{document}